\documentclass[journal,draftcls,onecolumn,12pt,twoside]{IEEEtranTCOM}

\IEEEoverridecommandlockouts
\usepackage{cite}
\usepackage{amsmath,amssymb,amsfonts}
\usepackage{algorithmic}
\usepackage{graphicx}
\usepackage{textcomp}
\usepackage{xcolor}
\usepackage{xspace}
\usepackage{cite}

\usepackage[nolist]{acronym}
\usepackage[draft,marginclue,footnote]{fixme}
\usepackage{subcaption}
\usepackage[
breaklinks=true
]{hyperref}
\usepackage{breakurl}
\usepackage{multirow}
\usepackage{rotating}
\usepackage{mathtools}
\usepackage{algorithmic}
\usepackage{algorithm}
\usepackage[section]{placeins}
\usepackage{cancel}
\usepackage{soul}

\usepackage{comment}

\let\Oldsection\section
\renewcommand{\section}{\FloatBarrier\Oldsection}

\usepackage{color}

\def\BibTeX{{\rm B\kern-.05em{\sc i\kern-.025em b}\kern-.08em
    T\kern-.1667em\lower.7ex\hbox{E}\kern-.125emX}}
\definecolor{darkgreen}{RGB}{0, 150, 0}
\definecolor{darkred}{RGB}{150, 0, 0}

\newcommand\undermat[2]{%
  \makebox[0pt][l]{$\smash{\underbrace{\phantom{%
    \begin{matrix}#2\end{matrix}}}_{\text{$#1$}}}$}#2}

\newcommand{\RealPlus}{\ensuremath{\mathbb{R^+}}\xspace} % set of positive real numbers
\newcommand{\N}{\ensuremath{\mathcal{N}}\xspace} % set of nodes
\newcommand{\C}{\ensuremath{\mathcal{C}}\xspace} % set of all cloud data centers
\newcommand{\F}{\ensuremath{\mathcal{F}}\xspace} % set of all fog nodes
\newcommand{\R}{ \ensuremath{R_k}\xspace} % set of all offloaded computational requests
\newcommand{\T}{\ensuremath{\mathcal{T}}\xspace} % length of evaluation period
\newcommand{\aki}{\ensuremath{a_{n}^{r}}\xspace} % whether request $k$ is computed at node $i$
\newcommand{\Ecpk}{\ensuremath{E_{\text{cp}}^{r}}\xspace} % energy spent on processing request $k \in \R$
\newcommand{\Ecpki}{\ensuremath{E_{\text{cp},n}^{r}}\xspace} 
\newcommand{\EcpkiTilde}{\ensuremath{\tilde{E}_{\text{cp},n}^{r}}\xspace} 
\newcommand{\EcpkiOptTilde}{\ensuremath{\tilde{E}{_{\text{cp},n}^{r}}}^\star\xspace} 

\newcommand{\EcpkiPlus}{\ensuremath{E{_{\text{cp},n}^{r+}}}\xspace} 
\newcommand{\EcpkiMinus}{\ensuremath{E{_{\text{cp},n}^{r-}}}\xspace} 
\newcommand{\EcpkiMinusTilde}{\ensuremath{\tilde{E}{_{\text{cp},n}^{r -}}}\xspace} 
\newcommand{\betai}{\ensuremath{\beta_{n}}\xspace} % energy efficiency (FLOPS per Watt) characterizing node $i \in \N$
\newcommand{\Lk}{\ensuremath{L^{r}}\xspace} % size of request $k \in \R$ 
\newcommand{\thetak}{\ensuremath{\theta^{r}}\xspace} % computational complexity of task $k \in \R$
\newcommand{\Ecommk}{\ensuremath{E_{\text{comm}}^{r}}\xspace} % energy spent on transmission of request $k \in \R$
\newcommand{\Ecommki}{\ensuremath{E_{\text{comm},n}^{r}}\xspace} 
\newcommand{\bk}{\ensuremath{g^{r}}\xspace} % \ac{FN} to which the request $k \in \R$ is originally sent
\newcommand{\ok}{\ensuremath{o^{r}}\xspace} % output-to-input data size ratio of request $k \in \R$
\newcommand{\gammaik}{\ensuremath{\gamma_n^{r}}\xspace} % energy-per-bit cost of transmitting data of request $k \in \R$ between \acp{FN} \bk  and $i$
\newcommand{\Etotk}{\ensuremath{E_{\text{tot}}^{r}}\xspace} % energy spent on transmission and processing of request $k \in \R$
\newcommand{\EtotkiOpt}{\ensuremath{E{_{\text{tot},n}^{r}}}^\star\xspace}
\newcommand{\EtotkiOptTilde}{\ensuremath{\tilde{E}{_{\text{tot},n}^{r}}}^\star\xspace}

\newcommand{\Etotki}{\ensuremath{E_{\text{tot},n}^{r}}\xspace}
\newcommand{\EtotkiTilde}{\ensuremath{\tilde{E}_{\text{tot},n}^{r}}\xspace}

\newcommand{\Dcpk}{\ensuremath{D_{\text{cp}}^{r}}\xspace} % delay caused by processing request $k \in \R$
\newcommand{\Dcpki}{\ensuremath{D_{\text{cp},n}^{r}}\xspace}
\newcommand{\freqi}{\ensuremath{f_n}\xspace} % clock frequency of node $i \in \N$
\newcommand{\freqidelay}{\ensuremath{f_{ \mathrm{delay}, n}^r}\xspace} % clock frequency of node $i \in \N$
\newcommand{\freqimin}{\ensuremath{f_{\text{min},n}}\xspace} % minimum clock frequency of node $i \in \N$
\newcommand{\freqimax}{\ensuremath{f_{\text{max},n}}\xspace} % maxium clock frequency of node $i \in \N$
\newcommand{\si}{\ensuremath{s_n}\xspace} % number of FLOPs performed per single clock cycle at node $i \in \N$
\newcommand{\Dcommki}{\ensuremath{D_{\text{comm},n}^{r}}\xspace} 
\newcommand{\ribk}{\ensuremath{b^{n}_{r}}\xspace} % link bitrate from \acp{FN} $i \in \F$ to $\bk \in \F$
\newcommand{\rback}{\ensuremath{b_{\text{back}}}\xspace} % link bitrate in the backhaul and backbone network
\newcommand{\di}{\ensuremath{d^{n}}\xspace} % fiberline distance to Cloud \ac{DC} $i$.
\newcommand{\chiRTT}{\ensuremath{\chi}\xspace} % parameter characterizing (in \textmu s/km) \ac{RTT} \cite{Almes1999, Olbrich2009} dependence on distance.
\newcommand{\Dqueueki}{\ensuremath{D_{\text{queue},n}^{r}}\xspace} 
\newcommand{\Dtotk}{\ensuremath{D{_{\text{tot}}^{r}}}\xspace} % total offloading delay of request $k \in \R$
\newcommand{\Dtotki}{\ensuremath{D_{\text{tot},n}^{r}}\xspace}
\newcommand{\Dmaxk}{\ensuremath{D_{\text{max}}^{r}}\xspace} % delay threshold for request $k \in \R$
\newcommand{\Pacti}{\ensuremath{P_{n}}\xspace} % power consumption of a device in active-state

\newcommand{\tk}{\ensuremath{T_k}\xspace} % time at which request k arrives
\newcommand{\ti}{\ensuremath{t_{n}}\xspace} % time at which FN i finishes computing last task 
\newcommand{\Dcommku}{\ensuremath{D_{\text{ul},n}^{r}}\xspace} % delay caused by transmitting request $k \in \R$ -- UL
\newcommand{\Dcommkd}{\ensuremath{D_{\text{dl},n}^{r}}\xspace} % delay caused by transmitting request $k \in \R$ -- DL
\newcommand{\powerParamThreeI}{\ensuremath{p_{n,3}}\xspace} % power model parameter
\newcommand{\powerParamTwoI}{\ensuremath{p_{n,2}}\xspace} % power model parameter
\newcommand{\powerParamOneI}{\ensuremath{p_{n,1}}\xspace} % power model parameter
\newcommand{\powerParamZeroI}{\ensuremath{p_{n,0}}\xspace} % power model parameter

\newcommand{\refFig}[1]{Fig.~\ref{#1}}
\newcommand{\refFigs}[1]{Figs.~\ref{#1}}

\newcommand{\refSec}[1]{Section~\ref{#1}}
\newcommand{\refTab}[1]{Table~\ref{#1}}
\newcommand{\refEq}[1]{Eq.~\eqref{#1}}

\newcommand{\refAlg}[1]{Alg.~\ref{#1}}

\newcommand{\etal}{\textit{et~al.} }
\newcommand{\ie}{i.e., }
\newcommand{\eg}{e.g., }
\newcommand{\Eg}{E.g., }
          
\graphicspath{{./figures/}{./plots/}}     

\begin{document}

\includecomment{onecolumn}
\excludecomment{twocolumn}

\title{Task Allocation for Energy Optimization in Fog Computing Networks with Latency Constraints
\thanks{This work is funded by the Ministry of Education and Science, Poland (subvention 0312/SBAD/8159).}}

\author{\IEEEauthorblockN{Bartosz Kopras,
Bartosz Bossy,
Filip Idzikowski,
Pawe\l{} Kryszkiewicz,
Hanna Bogucka}

\IEEEauthorblockA{
Poznan University of Technology, Poland}}

%\listoffixmes
\maketitle

%\vspace{-1cm}
\begin{abstract}
\textcolor{blue}{This work has been submitted to the IEEE for possible publication. Copyright may be transferred without notice, after which this version may no longer be accessible.}
	
Fog networks offer computing resources with varying capacities at different distances from end users.
A \ac{FN} closer to the network edge may have less powerful computing resources compared to the cloud, but processing of computational tasks in an \ac{FN} limits long-distance transmission.
How should the tasks be distributed between fog and cloud nodes? 
We formulate a universal non-convex \ac{MINLP} problem minimizing task transmission- and processing-related energy with delay constraints to answer this question.
It is transformed with \ac{SCA} and decomposed using the primal and dual decomposition techniques. 
Two practical algorithms called \ac{EEFFRA} and \ac{LC}-\ac{EEFFRA} are proposed.
They allow for successful distribution of network requests between \acp{FN} and the cloud in various scenarios significantly reducing the average energy cost and decreasing the number of computational requests with unmet delay requirements.

\end{abstract}


%\section*{List of Acronyms}
\begin{acronym}[FLOPS\_\_]
	\setlength{\parskip}{0ex}
	\setlength{\itemsep}{0.01ex}
	
	\acro{AWS}{Amazon Web Services}
	\acro{AP}{Access Point}
	\acro{API}{Application Programming Interface}
	\acro{BAN}{Body Area Network}
	\acro{BS}{Base Station}
	\acro{BSN}{Body Sensor Network}
	\acro{CDN}{Content Delivery Network}
	\acro{CDF}{Cumulative Distribution Function}
	\acro{CPE}{Customer Premises Equipment}
	\acro{CN}{Cloud Node}
	\acro{CPU}{Central Processing Unit}
	\acro{C-RAN}{Cloud Radio Access Network}
	\acro{DC}{Data Center}
	\acro{DVFS}{Dynamic Voltage and Frequency Scaling}
	\acro{EA}{Energy-Aware}
	\acro{ECG}{Electrocardiogram}                 
	\acro{EEFFRA}{Energy-EFFicient Resource Allocation}
	\acro{EH}{Energy Harvesting}
	\acro{ETSI}{European Telecommunications Standards Institute}
	\acro{EPON}{Ethernet Passive Optical Network}
	\acro{FI}{Fog Instance}
	\acro{FLOP}{Floating Point Operation}
	\acro{FLOPS}{Floating Point Operations per Second}
	\acro{FN}{Fog Node}
	\acro{F-RAN}{Fog Radio Access Network} 
	\acro{GPS}{Global Positioning System}
	\acro{GSM}{Global System for Mobile communications}
	\acro{HD}{High Definition}
	\acro{IBStC}{If Busy Send to Cloud}
	\acro{IBKiF}{If Busy Keep in FN}
	\acro{IBStOF}{If Busy Send to Other FN}
	\acro{IEEE}{Institute of Electrical and Electronics Engineers} 
	\acro{ICT}{Information and Communication Technology}
	\acro{IoE}{Internet of Everything}
	\acro{IoT}{Internet of Things}
	\acro{IP}{Internet Protocol}
	\acro{KKT}{Karush--Kuhn--Tucker}
	\acro{LAN}{Local Area Network}
	\acro{LC}{Low-Complexity}
	\acro{LTE}{Long Term Evolution}         
	\acro{MCC}{Mobile Cloud Computing}
	\acro{MINLP}{Mixed-Integer Nonlinear Programming}
	\acro{MD}{Mobile Device}
	\acro{MEC}{Mobile/Multi-Access Edge Computing}
	\acro{MIB}{Management Interface Base}   
	\acro{nDC}{nano Data Center}
	\acro{NFV}{Network Function Virtualization}
	\acro{OSI}{Open Systems Interconnection}
	\acro{PA}{Power-Aware}                   
	\acro{PC}{Personal Computer}             
	\acro{PGN}{Portable Game Notation} 
	\acro{QoE}{Quality of Experience}
	\acro{QoS}{Quality of Service}           
	\acro{RAM}{Random Access Memory}        
	\acro{RAN}{Radio Access Network}
	\acro{RFC}{Request for Comments}
	\acro{RRH}{Remote Radio Head} 
	\acro{RTT}{Round-Trip Time}
	\acro{SCA}{Successive Convex Approximation}
	\acro{SCN}{Small Cell Network}
	\acro{SDN}{Software Defined Network}
	\acro{SDR}{SemiDefinite Relaxation} 
	\acro{SINR}{Signal to Interference-plus-Noise Ratio}
	\acro{SM}{Sleep Mode}                    
	\acro{SOA}{Service Oriented Architecture}
	\acro{TDM}{Time Division Multiplexing}   
	\acro{TE}{Traffic Engineering}           
	\acro{TM}{Traffic Matrix}  
	\acrodefplural{TM}{Traffic Matrices}    
	\acro{V2V}{Vehicle-to-Vehicle}
	\acro{V2X}{Vehicle-to-Anything}
	\acro{VC}{Virtual Cluster}             
	\acro{VM}{Virtual Machine}
	\acro{WAN}{Wide Area Network}
	\acro{WDM}{Wavelength Division Multiplexing}
	\acro{WE}{WeekEnd day}                   
	\acro{WLAN}{Wireless Local Area Network}

\end{acronym}

\setlength\emergencystretch{\hsize} %prevents long words/equations from causing "overfull" lines
% !TEX root = main.tex

% ======================================================
%    INTRODUCTION
% ======================================================
\section{Introduction}
\label{sec:introduction}

\begin{figure}
	\centering
	\includegraphics[width=0.48\textwidth]{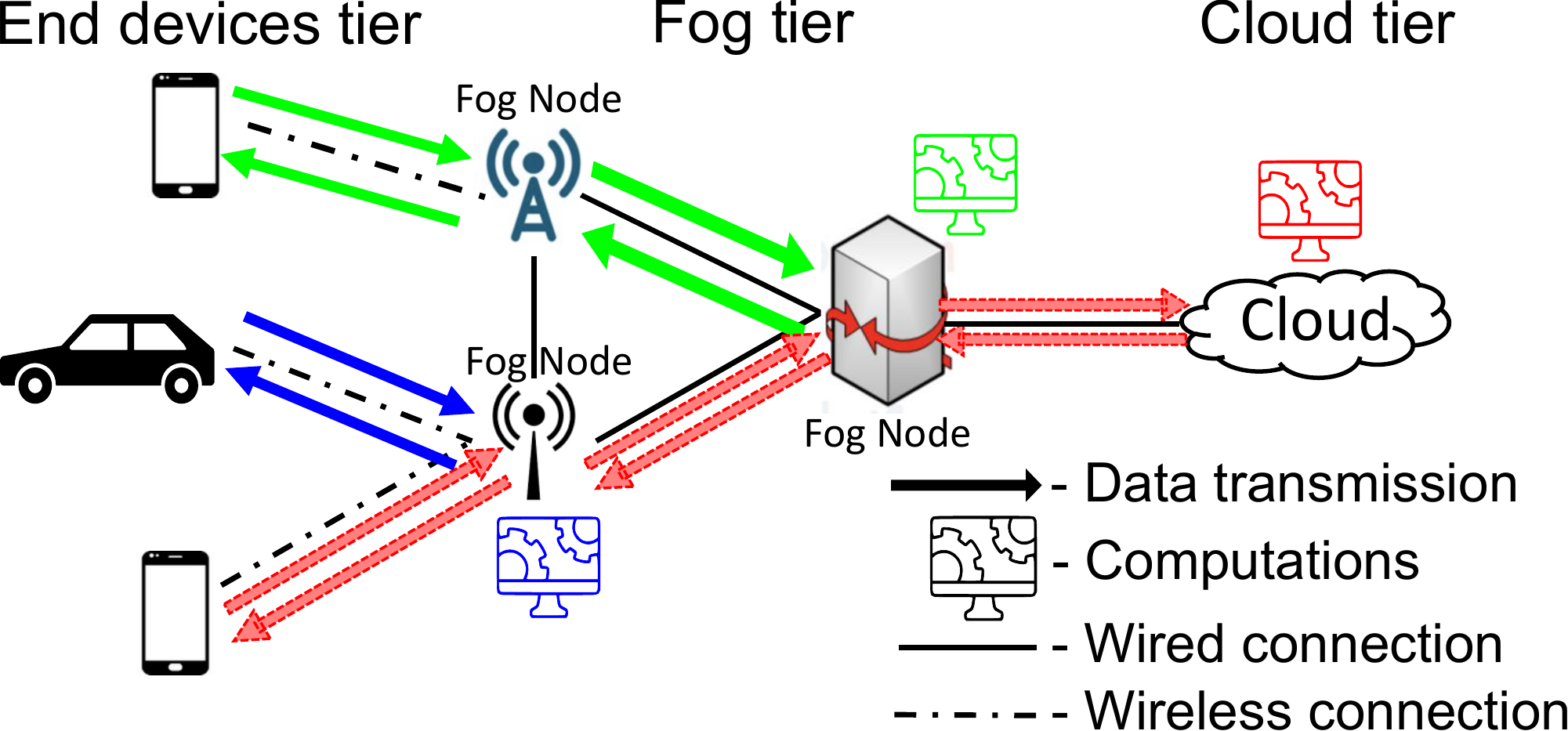}
	\caption{Fog computing network  with task offloading scenario (colors differentiate offloaded tasks).}% -- arrows represent flow of data and monitors represent computations.}
	\label{fig:fog_offloading}
	\vspace{-0.5cm}
\end{figure}

The number of \ac{IoT} devices continues to grow~\cite{Ericsson2018}. 
%This creates new challenges for the \ac{ICT} sector.
%\fmfi{Shall ICT be expanded in the previous sentence?}
They often have limited memory capacity  and computational power.
%, and battery life related to their small sizes.
%These limitations hinder their ability to process gathered data.
However, processing requests of all \ac{IoT} devices at a remote cloud would require an unprecedented amount of traffic traversing the Internet~\cite{Cisco2018} and influencing energy consumption as well as congestion of the Internet.
%which would influence the total energy consumed by the Internet.
Moreover, some applications such as video surveillance,
%~\cite{ Chen2016a},
augmented reality, 
%~\cite{Zao2014}
%industry and medical automation
or vehicle-to-vehicle communication
require low delays that cannot be fulfilled by remote cloud \acp{DC}.
%\fmfi{Potentially drop \cite{ Chen2016a} and \cite{Zao2014}.}
%Tasks performed for these applications are often too complex (and battery-draining) to be processed by \acp{MD}.
%Offloading these tasks to a Cloud \ac{DC} could introduce unacceptable delays.
%level of latency.
Fog computing \cite{Bonomi2012} addresses these problems by introducing a fog tier (or multiple hierarchical fog tiers) between the \emph{cloud} and the \emph{end devices} tiers (\refFig{fig:fog_offloading})\footnote{ \emph{Fog-computing} is closely related to \emph{edge-computing}, but we distinguish \emph{the fog} from \emph{the edge} by the mentioned multiple hierarchical layers of the fog and flexibility of directing the computational tasks to suitable \acp{FN} introduced by such architecture.}.
It is composed of \acfp{FN} with computational and storage resources located near end users.
%``on the edge'' of the network.
Data transmission between \emph{end devices} and \acp{FN} is thus faster and potentially less energy-consuming than alternative thing-to-cloud communication.

This work considers task distribution between many \acp{FN} and \acp{CN}.
We aim at minimization of network energy consumption while meeting delay constraints specific for each offloaded task. As seen in \refFig{fig:fog_offloading} an offloaded task can be processed in the node to which it is originally sent (solid blue arrows), in another \ac{FN} (solid green arrows), or in the cloud (hollow red arrows).
%\fmfi{@BK: Why cannot the arrows be solid, dotted and dashed in \refFig{fig:fog_offloading}? BK: added source file: figures Fog\_offloading\_corrected3.svg }
A realistic network model is proposed below encompassing the energy consumption as well as the task-implementation delay, related to both the necessary computations and communication.
Our model includes realistic network parameters reflecting characteristics of real-world equipment.
Based on this model, we formulate the optimization problem to minimize the total (task transmission- and implementation-related) energy consumption while fulfilling total delay constraints.
The optimization considers not only the assignment of tasks to the nodes %, 
but also the \ac{CPU} frequency at each utilized node.
The problem is a non-convex \acf{MINLP} problem, so we apply the \acf{SCA} method which transforms it to a series of convex \ac{MINLP} problems, and provide the optimal solution by using the primal and dual decomposition techniques and the Hungarian algorithm.
A sub-optimal, lower complexity solution %which does not require coordinated decision making 
is also proposed.
The main contributions of our work include: 1. modeling energy consumption and delays related to transmission, queuing, and computing of tasks in the fog and cloud tiers, 2. proposing and solving a complex optimization problem, 3. examining  the efficiency of proposed solutions for a vast scope of network and traffic parameters. 
%{\HB This optimal solution however, has extremely high computational cost that} can be  prohibitive in a real network {\HB scenario. Therefore, in this paper, we explore} suboptimal solutions {\HB to this problem and provide low-complex solutions, which are then evaluated against the optimal one.} %NOT TRUE
Our solutions are original in handling more complex network models, more realistic parameters, and considering both task processing- (computing-) and transmission-related energy and delay as opposed to the former and background works.
%{\BK rewrite last sentence, "main contributions of the paper are:"}

The paper is organized as follows.
The overview of related work is provided in \refSec{sec:related-work}.
The mathematical model of the fog network is presented in \refSec{sec:problem}.
\refSec{sec:opt} defines the optimization problem, and \refSec{sec:solution} presents its solution.
Simulation results for various scenarios are shown and discussed in \refSec{sec:results}.
\refSec{sec:conclusion} concludes~the~work.
%{\PK Ogolnie pytanei co wyrzucamy z tego artykulu? Moze te suboptimal solutions? Ale wtedy trzeba miec dostep do plikow fig wynikow.}
%{\BK usuniecie ich z wykresow nie byloby trudne, ale za duzo na tym nie zyskamy raczej. Wykres zajmuje tyle samo miejsca niezaleznie od liczby linii. \refSec{sec:low-compl-sol} to 1/6 strony.}

% ======================================================
\begin{comment}
\begin{figure}
	\centering
	\includegraphics[width=0.48\textwidth]{Fog_tasks123.pdf}
	\caption{ Tasks performed by Fog computing network. }
		%{\PK Wlasciwie to ten rysunek jest bardzo podobny do poprzedniego. Lepiej byloby go przerysowac tak ,zeby uwzgledniam nasza notacje: r, o, n ty wezel itp albo nie umieszczac}
%	\fmbk{@BK: Leave just offloading... or not}
	\label{fig:fog}
\end{figure}
\end{comment}
% ======================================================

%\subsection{Purpose of Fog Computing}
%Three broad types of services provided by Fog network are shown in %\refFig{fig:fog}.
%They include (1) offloading, (2) content distribution, and (3) data aggregation.

%These concepts are briefly discussed below.
%
%
%Let us start with offloading as that is the concept most often associated with Fog.

% !TEX root = main.tex
\section{Related Work}
\label{sec:related-work}
{
 We review related work while focusing on the following aspects:
(i) energy consumption of fog networks related to tasks implementation (computations) and %to 
transmission of computational tasks,
(ii) consideration of latency %of computation tasks with delays
caused by both computations and transmission,
(iii) fog scenarios, network and traffic models, 
%(iv) methods for energy saving.
(iv) optimization of energy consumption.
The aspects (i)--(iv) stress the novelty of our work,
where we jointly study communication and computing in the fog in the context of offloading of computational tasks.
Works which focus on other potential applications of fog computing 
(\eg content distribution/caching \cite{Chen2016,Xu2018,Wang2020}) are therefore not included in this overview as their network and traffic scenarios 
are significantly different from ours.
Papers not taking energy consumption into account, 
 \eg  \cite{Liu2020,Hussein2020,Phan2021} and those who only look at power consumption as a constraint,  \eg  \cite{Lin2020} 
are also intentionally omitted.

The authors of \cite{Dinh2017,You2017,Liu2018} look at the energy consumption of individual end devices.
They answer the following question: is it more efficient for an end-device to process a task locally, or to offload it to the fog/cloud?
These works disregard the energy costs related to performing computations and transmission in the fog or cloud tiers of the network (our aspect (i)) %, 
and are therefore substantially different from our work.
Our work and those we survey further below examine  total energy consumption from the network operator point of view,
tackling the following question: given a task has already been offloaded, where is it favorable to process it?

Sarkar \etal examine the energy consumption \cite{Sarkar2016}, the power consumption \cite{Sarkar2018}, and the task delay \cite{Sarkar2016,Sarkar2018} in the fog-computing network, depending on how much data is processed in the fog tier and in the cloud tier.
This data is transmitted from terminal nodes.
Their models include costs related to transmission, processing, and storage of data.
These works show that both power/energy consumption and delay decrease with a higher percentage of tasks being processed in \acp{FN} rather than the cloud.
Power/energy consumption and delay also increase with a number of terminal nodes. 
However, this increase is not monotonous.
Moreover, a discussion on application-specific fog computing utilization is included in \cite{Sarkar2016}, while costs related to the carbon footprint of the network are examined in \cite{Sarkar2018}.

There are multiple major differences between our work and \cite{Sarkar2016,Sarkar2018}.
First, there is no optimization (aspect (iv)) in these works.
While costs are modeled depending on where offloaded requests are sent, no solution for their optimization is proposed. 
Second, \acp{FN} in \cite{Sarkar2016,Sarkar2018} work at fixed clock-frequency while in our work, \ac{DVFS} is considered (aspect (iii)).
Furthermore, costs related to transmission within the fog tier (aspects (i) and (ii)) are also not considered in \cite{Sarkar2016,Sarkar2018}.
Finally, peculiar assumptions (indefinite processing of data in cloud \acp{DC} \cite{Sarkar2018}, { energy dissipation rate defined as a sum of energy spent over time on computations plus an average of energy spent over time on transmission \cite{Sarkar2016}}) and mistakes (\eg triangle inequality used incorrectly in Sec. 5.2.1 of \cite{Sarkar2018}) make results of \cite{Sarkar2016, Sarkar2018} biased towards showing that fog computing is significantly faster and more energy-efficient than cloud computing regardless of the offloading scenario.
%Moreover, what differs \cite{Sarkar2016,Sarkar2018}

In \cite{Deng2016}, offloaded requests can be served by one of the \acp{FN} or the cloud servers.
%The authors of \cite{Deng2016} formulate an optimization problem minimizing weighted sum of power consumption and delay.
Similar to our work, the objective of \cite{Deng2016} is to minimize the power consumption of the network while maintaining delay constraints.
 Delay related to queuing and computing is calculated as an average response time of queuing models %(with one serving node in \acp{FN} and multiple serving nodes in \acp{DC})
(M/M/1 in \acp{FN} and M/M/n in \acp{DC}).
%Optimization 
The cloud servers can adjust their clock frequency using \ac{DVFS}.
Results show a clear trade-off between power consumption and delay.
The aspects (i), (iii), and (iv) distinguish \cite{Deng2016} from our work.
The energy/power costs are not related to data transmission in \cite{Deng2016} (i).
%Transmitting data between nodes introduces delay, but no other costs in \cite{Deng2016}.
Also, rather than jointly optimizing energy costs related to tasks transmission and processing (computations) in the fog and %tasks implementation 
%in 
the cloud, Deng \etal \cite{Deng2016} heuristically split this optimization problem into three sub-problems (iv).
%Moreover, all offloaded traffic is parameterized with a single number (the workload) in \cite{Deng2016}.
Offloaded traffic is not divided into a number of requests, packets, or instances in \cite{Deng2016} (iii) and is only parameterized with a single number.
As a consequence, only the average delay can be calculated (and  constraint satisfied), rather than the individual delay of each offloaded request. 
In this work, traffic offloaded from end devices consists of multiple requests, where each request is defined with its size, arithmetic intensity, and delay requirement.

Vakilian \etal \cite{Vakilian2020,Vakilian2021} jointly optimize delay and energy consumption related to offloading in fog networks with multiple \acp{FN}.
\acp{FN} can cooperate by sending workload to each other and %to 
the cloud. 
Similarly to \cite{Deng2016}, delay related to queuing and computing is calculated as an average response time of queuing model (M/M/1 for workload processed in \acp{FN}) with a constant value added for transmission delay between \acp{FN}. %The differences between \cite{Vakilian2020} and \cite{Vakilian2021} are two-fold: 
While both works use similar objective functions (weighted sum of energy consumption and delay), \cite{Vakilian2021} includes \emph{fairness coefficients} that depend on resources available to \acp{FN} while \cite{Vakilian2020} does not. 
The authors conclude that the problem in \cite{Vakilian2020} is convex.
For \cite{Vakilian2021}, they propose a heuristic, population-based algorithm, \ie cuckoo evolution algorithm.
Results of \cite{Vakilian2020} show a clear trade-off between energy consumption and delay in the network.
The method proposed in \cite{Vakilian2021} decreases both delay and energy consumption compared with a competing algorithm { (proposed in \cite{Yifan2019})}, while both works highlight that cooperation between nodes leads to lowered costs.

Vakilian \etal \cite{Vakilian2020,Vakilian2021} do not consider adjusting \ac{CPU}  frequencies of nodes.
%which differs them from our work in aspect (iv).
Furthermore, offloaded traffic is not divided into a number of requests, packets, or instances in \cite{Vakilian2020,Vakilian2021}.
It is only parameterized with a single number (like in \cite{Deng2016})
Hence, \cite{Vakilian2020,Vakilian2021} differ from our work in aspect (iii)).
Finally, joint energy and latency minimization is performed in \cite{Vakilian2020,Vakilian2021}, while energy minimization under latency constraints is performed in our work -- aspect (iv).

Cai \etal \cite{ Cai2020} examine a network with a single task node and multiple helper nodes which lack an external power supply.
The task node can process computational tasks itself or offload these tasks to the helper nodes.
To do so, it needs to transfer both the task and energy required for computations to the helper node.
The authors of \cite{ Cai2020} optimize cost defined as a weighted sum of delay and energy consumed by the task node for computations, energy transmission, and task transmission.
The constraints include a maximum task execution delay which cannot be exceeded and offloading enough energy for computations in helper nodes.
Optimization algorithms are proposed for scenarios with and without the possibility of queuing tasks.

There are multiple differences between \cite{ Cai2020} and our work.
In terms of energy consumption (i) our work minimizes energy consumed by all computing nodes while \cite{Cai2020} minimizes energy spent only by the task node, energy consumption of helper nodes is examined but only as a constraint. 
There are significant differences in network and traffic models (iii).
\cite{Cai2020} examines an energy harvesting scheme where helping nodes are gated by the amount of energy they receive.
In \cite{Cai2020} tasks originate from one task node while in our work they arrive from end devices at any of our \acp{FN}.
Different to our work, all nodes work at fixed clock frequencies without any connection to the cloud in \cite{Cai2020}.

Power consumption and delay in fog and cloud tiers of fog computing networks are studied in \cite{Kopras2019}.
Results are shown for various traffic and network parameters.
A trade-off between power consumption and delay is shown both in the number of \acp{FN} and their clock frequency (more \acp{FN} and higher frequencies mean higher power consumption and lower delay).
However, in \cite{Kopras2019}, similarly to \cite{Sarkar2018} and  \cite{Sarkar2016}, power consumption and delay in the fog computing network is only examined -- there is no energy-cost optimization (aspect (iv)).
Also, while delay and energy costs related to transmission between fog and cloud tiers are considered in \cite{Kopras2019}, the cost related to transmission within the fog tier is not considered (aspects (i) and (ii)).

To summarize, our work aims at minimization of energy consumed by both  networking and computing equipment in the fog computing network.
To the best of our knowledge, no prior work tackles the problem of jointly minimizing energy spent on both transmission and computation by distributing tasks between the fog and the cloud, while satisfying latency constraints and dynamically choosing optimal \ac{CPU} clock frequencies.
Finally, we take the energy- and latency-costs of inter-\ac{FN} communication into account in our optimization algorithms, while in the surveyed papers these are ignored or assumed to be negligible.

}

% !TEX root = main.tex
\section{Network model}
\label{sec:problem}
We introduce the network model in this section.
Notation is presented in \refTab{tab:notation}. 
Letters in superscript are used throughout this work as upper indices, not exponents, \eg \Lk does \textbf{not} denote $L$ to the power of $r$.

In the bottom tier of the network, there are end-devices (\eg smartphones, sensors) with some specific computational tasks.
We assume that serving these tasks requires offloading them, \ie they either cannot be processed in the end device or the end device chooses to offload them rather than process them locally.
Then, they can be processed either in the fog tier, consisting of set \F of \acp{FN},
or in the cloud tier (set \C of \acp{DC}).
A set of all computing nodes in the network is denoted as $\N = \F \cup \C$.

Unlike works that focus on end devices \cite{Dinh2017,You2017,Liu2018}, we examine energy consumption from the point of view of the fog network.
Modeling and optimizing wireless transmission is a key part of these works, e.g., allocating sub-channels to mobile devices in \cite{You2017} or interference affecting transmission rates in \cite{Liu2018}.
Our system model is agnostic to the wireless technology and model/condition of the channel used for transmission between end devices and \acp{FN}.
We examine requests as they appear in the fog tier of the network.
Instead, we focus on the efficient distribution of offloaded tasks between fog and cloud nodes.

\vspace{-0.5cm}

\subsection{Computational Requests}

Let 
\T be a numbered set $\{T_1, T_2, ... ,T_{|\T|} \}$ of all time  instances at which computational requests  %described by the set \R 
arrive in the network, and have to be allocated computing resources. 
Let 
$\R$ be a set of all requests arriving in the network at time $T_k$.
Each computational request $r \in \R$ is described by the following parameters:
    (i) size \Lk in bits,
    (ii) 
     arithmetic intensity %\cite{Zhao2018,Allayla2021}
    \thetak  in FLOP/bit (used in FLOP/byte in \cite{Zhao2018,Allayla2021}),
    (iii) ratio \ok of the size of the result to the size of the offloaded task {(most related works do not consider the transmission of the results \cite{Sarkar2016,Sarkar2018} or assume that its contribution is negligible \cite{Deng2016,Cai2020}; \ok equal to 1 implies the output has the same size as the input; in case of electrocardiography signals $\ok \simeq 0.07$ \cite{Gia2015})},
    (iv) \ac{FN} $\bk \in \F$ to which the request is originally sent (before allocation) 
   
    (v) maximum tolerated delay \Dmaxk.
Let us define a binary variable \aki that shows %whether request $r \in \R$ is computed at node $n \in \N$.
 where the request is computed, i.e., \aki equals 1 if $r \in \R$ is computed at node $n \in \N$, and 0 otherwise.

\vspace{-0.5cm}

\begin{table}
	\renewcommand{\arraystretch}{0.7}
	\caption{Notation for modelling fog computing network and defining optimization problem.%{\BK @All: brakuje notacji z Sekcji IV}
	}
	\label{tab:notation}
	%\vspace{-2mm}
	\centering
	\begin{tabular}{c p{1.0cm} p{12.0cm}}

		\hline
		& Symbol & Description \\
		\hline

		\multirow{21}{*}{\begin{sideways}Parameters\end{sideways}}
&	 \rback 	&	 link bitrate in the backhaul and backbone network \\
&	 \ribk 	&	 link bitrate between \acp{FN} $n \in \F$ and $\bk \in \F$ \\
&	 \C 	&	 set of all cloud \acp{DC} \\
&	 \chiRTT 	&	 a parameter characterizing delay depending on distance\\
&	 \di 	&	 fiberline distance from the fog to cloud \ac{DC} $n \in \C$ \\
&	 \Dmaxk 	&	 maximum tolerated delay requirement for request $r \in \R$ \\
&	 \F 	&	 set of all Fog Nodes \\
&	 \freqimax 	&	 maximum clock frequency of node $n \in \N$ \\
&	 \freqimin 	&	 minimum clock frequency of node $n \in \N$ \\
	&	 \bk 	&	 \ac{FN} to which the request $r \in \R$ is originally sent \\

&	 \gammaik 	&	 energy-per-bit cost of transmitting data of request $r \in \R$ between nodes $\bk \in \F$  and $n \in \N$ \\
&	 \Lk 	&	 size of request $r \in \R$ \\
&	 \N 	&	 set of all nodes \\
&	 \ok 	&	 output-to-input data size ratio of request $r \in \R$ \\
&	 $p_{n,q}$
&	  $q$-th coefficient of polynomial modeling power consumption of \ac{CPU} installed in node $n \in \N$ \\
&	 $Q$ 	&	 degree of polynomial \Pacti\\
&	 \R 	&	 { set of all computational requests offloaded at \tk }\\
&	 $\R'$ 	&	 { set of rejected computational requests offloaded at \tk }\\
&	 \si 	&	 number of FLOPs performed per single clock cycle at node $n \in \N$ \\
&	 \T 	&	 set of all considered time instances, when one or more computational requests arrive \\
&	 \thetak 	&	 arithmetic intensity of request $r \in \R$ \\
&	 \tk 	&	 time at which request $r \in \R$ arrives in the network, $k \in \{1, ..., |\T|\}$ \\
&	 \ti 	&	 time at which node $n \in \N$ finishes computing its last task\\

		\hline
		\multirow{21}{*}{\begin{sideways}Variables and metrics\end{sideways}}
		&	 \aki 	&	 whether request $r \in \R$ is computed at node $n \in \N$, $\aki \in \{0, 1\}$  \\
&	 \betai 	&	 energy efficiency (FLOPS per Watt) characterizing node $n \in \N$ \\
&	 \Dcommki 	&	 delay caused by transmitting request $r \in \R$ between nodes $\bk \in \F$ and $n \in \N$ \\
&	 \Dcommkd 	&	 delay caused by transmitting request $r \in \R$ between nodes $\bk \in \F$ and $n \in \N$ -- downlink \\
&	 \Dcommku 	&	 delay caused by transmitting request $r \in \R$ between nodes $\bk \in \F$ and $n \in \N$ -- uplink \\
&	 \Dcpk 	&	 computational delay caused by processing request $r \in \R$ in the network\\
&	 \Dcpki 	&	 computational delay caused by processing request $r \in \R$ at node $n \in \N$ \\
&	 \Dmaxk 	&	  maximum tolerated delay requirement for request $r \in \R$ \\
&	 \Dqueueki 	&	 queuing  delay of request $r \in \R$ at node $n \in \N$ \\
&	 \Dtotk 	&	 total delay of request $r \in \R$ \\
&	 \Dtotki 	&	 total delay of processing request $r \in \R$ at node $n \in \N$ \\
&	 \Ecommk 	&	 energy spent on transmission of request $r \in \R$ \\
&	 \Ecommki 	&	 energy cost for transmission of request $r \in \R$ between nodes $\bk \in \F$ and $n \in \N$\\
&	 \Ecpk 	&	 energy spent in the network on processing request $r \in \R$ \\
&	 \Ecpki 	&	 energy cost of processing request $r \in \R$ at node $n \in \N$ \\
&	 \Etotk 	&	 energy spent on transmission and processing of request $r \in \R$ \\
&	 \Etotki 	&	 energy cost of offloading request $r \in \R$ when computing it at node $n \in \N$ \\
&	 \freqi 	&	 clock frequency of node $n \in \N$%, 
\\
&	 \Pacti 	&	 power consumption related to computations at node $n \in \N$ \\

		\hline

	\end{tabular}
	
\end{table}

\subsection{Energy Consumption}
The energy consumption model consists of two parts: communication (transmission of data) and computation (processing of data). Energy \Ecpk spent on processing request $r \in \R$ equals:
\begin{equation}
\label{eq:power_cp}
\Ecpk=
\sum_{n \in \N}\aki
\Ecpki = \sum_{n \in \N}\aki \frac{\Lk\thetak}{\betai} \text{,}
\end{equation}
where  \Ecpki is the energy spent on processing of request $r \in \R$ at node $n \in \N$.
\betai characterizes computational efficiency of node $n \in \N$ given in { \acp{FLOP} per second} per watt \cite{Green2020}. 
For cloud \acp{DC}, we assume constant \ac{CPU} clock frequency \freqi and efficiency \betai.
For \acp{FN}, 
\betai depends on \ac{CPU} clock frequency \freqi of node $n \in \F$, its power consumption $\Pacti$, and number \si of \acp{FLOP} performed within a single clock cycle of this node \cite{Dolbeau2018}:
% ======================================================
\begin{equation}
\label{eq:beta3}
\betai=\frac{\freqi\si}{\Pacti} = \frac{\freqi\si} {\sum_{q=0}^{Q}p_{n,q}f_{n}^{q}} \text{.}
\end{equation}
% ======================================================

We model \Pacti as a $Q$-th degree polynomial of \freqi using parameters $p_{n,q}$ 
based on \cite{Park2013}.
This allows the model to cover various models of \acp{CPU}. Moreover, clock frequency \freqi must be within the range of minimum and maximum frequencies of the \ac{CPU} installed in node $n \in \F$, i.e., $\freqimin  \leq \freqi \leq \freqimax$.

The energy spent on the transmission of request  $r \in \R$ equals:
\begin{equation}
\Ecommk=
\sum_{n \in \N}\aki\Ecommki = \sum_{n \in \N}\aki \Lk(1+\ok)\gammaik \text{,}
\end{equation}
where \Ecommki is the energy required to transmit (communicate) request $r \in \R$ between \ac{FN} \bk and node $n\in\N$ while \gammaik is the energy-per-bit cost of transmitting data request $r \in \R$ between node $n$ and node \bk.
$\Lk\ok$ is the size (in bits) of results transmitted back to \ac{FN} \bk.
Thus, the total energy spent on offloading request $r \in \R$ 
is given by:
\begin{equation}
\label{eq:etotk}
\Etotk= \sum_{n \in \N} \aki \Etotki  = \sum_{n \in \N} \aki \left( \Ecpki+\Ecommki \right) \text{,}
\end{equation}
where \Etotki is the energy cost of offloading request $r \in \R$ { when it is computed at} node $n \in \N$.
{ Energy spent on wireless transmission between end devices and \acp{FN} is not included in \Etotki as we examine requests already sent by the end devices (as they appear in the \acp{FN}).}

\vspace{-0.5cm}
\subsection{Delay}
The delay model is %(similarly to the energy consumption model) 
divided into three parts: communication, computation, and queuing.
The delay \Dcpk caused by computing request $r \in \R$ equals:
\begin{equation}
\label{eq:Dcpk}
\Dcpk=
\sum_{n \in \N}\aki
\Dcpki =\sum_{n \in \N} \aki \frac{\Lk\thetak}{\freqi \si} \text{,}
\end{equation}
where \Dcpki is the time required to compute request $r \in \R$ at node $n \in \N$.
Moreover, there are significant differences between models of communication delay for requests processed in the fog tier and the cloud tier of the network.
It stems from the fact that clouds are assumed to have huge
(practically infinite)
computational resources with parallel-computing capabilities, and there is no need for queuing multiple requests served by the cloud \ac{DC} $n\in \C$.
They can be processed simultaneously.
Meanwhile, if multiple requests are sent to the same \ac{FN} $n\in \F$ for processing in a short time span, additional delays may occur due to congestion of computational requests (an arriving request cannot be processed until processing of all the previous requests has been completed).
On the other hand, it is assumed that cloud \acp{DC} are located far away from the rest of the network (hundreds or even thousands of kilometers away) which introduces additional, transmission-related  delay.
Delay caused by transmitting request $r \in \R$ between (to and from) \ac{FN} $\bk \in \F$ and cloud node $n \in \C$ is equal:
\begin{equation}
\label{eq:delay_cloud}
\Dcommki=
\frac{\Lk(1+\ok)}{\rback}
+\di\cdot\chiRTT,
\end{equation}
where \rback is the link bitrate in the backhaul and backbone network, while \di is the fiberline distance to cloud \ac{DC} $n\in \C$.
The parameter \chiRTT indicates the rate at which delay increases with distance \di \cite{Olbrich2009}.

For describing delays related to transmission between \acp{FN} let us split it into the uplink (sending  a request to be processed) and downlink (sending calculated results back to the origin of said request) parts denoted \Dcommku and \Dcommkd respectively.
For transmission between \acp{FN}, we assume delay caused by the distance between them ($\di\cdot\chiRTT$ in \refEq{eq:delay_cloud}) to be negligible~-- well below 1~ms as we use the value of $7.5\mu\text{s/km}$ for parameter \chiRTT \cite{Olbrich2009}~-- and therefore  we ignore it. The total delay caused by communication between \ac{FN} $\bk \in \F$ and $n \in \F$ for request $r \in \R$ equals:
\begin{equation}
\Dcommki=\Dcommku+\Dcommkd = \frac{\Lk}{\ribk} + \frac{\Lk\ok}{\ribk}\text{,}
\end{equation}
where \ribk is the link bitrate between \ac{FN} \bk and $n$.
As discussed earlier, when a request is sent to \ac{FN} $n\in \F$ and there is another request being processed at this node, the request is put in a queue and waits to be processed.
Let us define a scheduling variable $\ti \in \RealPlus$ which indicates when the processing of the last request scheduled at \ac{FN} $n \in \F$ is completed.

Queuing delay of the request $r \in \R$ 
at node $n \in \F$ is calculated as follows:
\begin{equation}
\label{eq:queue}
\Dqueueki=
\max(0,\ti-\tk-\Dcommku).
\end{equation}
\Dqueueki has positive values when $\ti > \tk+\Dcommku$, \ie when request $r$ arrives at node $n$ at time $\tk+\Dcommku$ and it is queued until processing of other request(s) is completed at time \ti. 
For each node $n \in \C$ (cloud \acp{DC}), \Dqueueki is always equal to zero, \ie each request arriving at the cloud can immediately be processed regardless of the number of requests already being processed due to parallel processing.
Thus, the total  delay of processing request $r\in \R$ 
is the sum of delays related to transmission, queueing, and computation:
\begin{equation}
\label{eq:Dtotk}
%\Dtotk=\Dcommku+\Dqueuek+\Dcpk+\Dcommkd
\Dtotk=\sum_{n \in \N}\aki\Dtotki= \sum_{n \in \N}\aki \left(\Dcommki+\Dqueueki+\Dcpki\right).
\end{equation}

\vspace{-0.1cm}
\subsection{Updating Scheduling Variables in the Fog}

Let us now explain how the values of scheduling variables \ti are assigned to become parameters of an optimization instance.
As no requests are processed at the beginning of the simulation, we set $\ti=0,\forall{n \in {\F}}.$
For each $\tk \in \T$, 
after allocations \aki are determined, times \ti are updated according to when computation of requests offloaded to \acp{FN} is scheduled to finish:
\begin{equation}
\label{eq:ti}
\ti:=\max(\ti,\tk+\sum_{r \in \R}\aki
(\Dcommku+\Dqueueki+\Dcpki)),\forall{n \in\F}.
\end{equation}
 By using \eqref{eq:ti}, each new instance of the optimization problem depends on results (allocations) of previous instances. 

\section{Optimization Problem}
\label{sec:opt}
The objective of our formulated problem is to minimize total energy spent on offloading all requests arriving at the network at time $T_k$, that is to find: 
 \begin{equation}
 \label{eq:obj_function}
 \left(\mathbf{a}^\star,\mathbf{f}^\star \right) = 
 \arg \min_{\mathbf{a},\mathbf{f}} \sum_{r \in \R} \Etotk \text{,}
\end{equation}
subject to:
\begin{align}
\label{eq:c1} 
&\quad \sum_{n \in \N} \aki = 1&  &\forall{r \in \R} \text{,}  \\
\label{eq:c2}
&\quad \sum_{r \in \R} \aki \leq 1, & &\forall{n \in  \F } \text{,} \\
\label{eq:delayConstraint}
&\quad \Dtotk \leq \Dmaxk, & & \forall{r \in \R} \text{,} \\
\label{eq:fminfmax}
&\quad \freqimin  \leq \freqi \leq \freqimax & &\forall{n \in \F} \text{,} \\
\label{eq:akiIn01}
&\quad \aki \in \{0,1\}, & & \forall{r \in \R} \text{,}\:  \forall{n \in \N} \text{,} 
\end{align}
where $\mathbf{a^\star} = \left\{ a{_n^r}^\star \right\}$ and  $\mathbf{f^\star} = \left\{ f_n^\star \right\}$ are optimal values of the optimization variables: allocation variables \aki and \ac{CPU} clock frequencies \freqi, respectively.
The constraints \eqref{eq:c1} and \eqref{eq:c2} restrict that each request must be processed at one and only one \ac{FN} or cloud \ac{DC} and that each \ac{FN} can process at most a~single request at a time, respectively.
The constraints \eqref{eq:delayConstraint} guarantee that the total delay \Dtotk must not be greater than the maximum acceptable one \Dmaxk.
Moreover, according to the constraints \eqref{eq:fminfmax}, the CPU frequency is limited by lower and upper bound while the decision variables \aki take only binary values, according to constraints \eqref{eq:akiIn01}.

The optimization problem cannot be solved for some sets of requests $\R$, where it is impossible to satisfy all constraints (\eg no feasible allocation of requests so that each request is processed  \eqref{eq:c1} while fulfilling its delay requirement \eqref{eq:delayConstraint}). 
In this case, rather than terminating the optimization without any solution (which would translate to rejecting all requests $\R$), we choose to reject requests for which \eqref{eq:delayConstraint} cannot be fulfilled.
The optimization is then performed over the set of remaining requests $\R \setminus \R'$, where $\R'$ denotes the set of rejected requests.

\section{Proposed solution}
\label{sec:solution}

The optimization problem defined in \refSec{sec:opt} is a \ac{MINLP} problem due to binary values of the allocation variables and continuous values of the \ac{CPU} clock frequencies. % while the 
Nonlinearity in the problem results from the power consumption model of the \ac{CPU} and the set of constraints \eqref{eq:delayConstraint}. 
Note that after substituting \eqref{eq:beta3} into \eqref{eq:power_cp}, the energy spent on processing of request $r \in \R$ at node $n \in \F$ is the sum of polynomial and rational functions:
\begin{align}
\label{eq:Ecp_parameters}
    \Ecpki = \frac{\Lk \thetak}{\si} \left[
    \frac{\powerParamZeroI}{\freqi} + \sum_{q=1}^{Q}p_{n,q}f^{q-1}
    \right] \text{.}
\end{align}
As such, for $\freqi \in \RealPlus$, the convexity of \eqref{eq:Ecp_parameters} in $\freqi$ depends on the parameters $p_{n,q}$ {(except $p_{n,1}$ and $p_{n,2}$ which have no influence on convexity, since their second derivatives are zero%they become zero in the second derivative
)}. If $\{ p_{n,0},p_{n,3},...,p_{n,Q}\}$
are positive, the objective function is convex.
If { all} these parameters are negative the function is concave. 
In these cases, standard optimization methods can be used to solve it \cite{boyd}.
However, if some of these parameters are negative, the others being positive, we deal with the difference of convex functions which is non-convex, requiring special optimization techniques. 
Therefore, in this section, the solution of the optimization problem, in the case of any possible values of CPU power consumption parameters
is presented as follows. 

Let us rewrite the objective function \eqref{eq:obj_function}
with \Ecpki being a difference of convex functions:
\begin{twocolumn}
\begin{align}
\label{eq:objFunDC}
    &  \left(\mathbf{a}^\star,\mathbf{f}^\star \right) =
    \\ \nonumber
    & \quad \arg \min_{\mathbf{a},\mathbf{f}} \sum_{r \in \R} \sum_{n \in \F} \aki  \left(\underbrace{\EcpkiPlus - \EcpkiMinus}_{\Ecpki} + \Ecommki \right)  \text{,}
\end{align}
\end{twocolumn}
\begin{onecolumn}
\begin{align}
\label{eq:objFunDC}
    &  \left(\mathbf{a}^\star,\mathbf{f}^\star \right) =  \arg \min_{\mathbf{a},\mathbf{f}} \sum_{r \in \R} \sum_{n \in \F} \aki  \left(\underbrace{\EcpkiPlus - \EcpkiMinus}_{\Ecpki} + \Ecommki \right)  \text{,}
\end{align}
\end{onecolumn}
 where \EcpkiPlus is the sum components of \Ecpki with positive parameters $p_{n,q}$ and \EcpkiMinus is the negative of the sum components of \Ecpki with negative parameters $p_{n,q}$.
We apply the \ac{SCA} method \cite{Bossy2020,Bossy_Globecom_2018,Wang2012} to approximate the possibly non-convex function by the series of convex ones.
Since the objective function \eqref{eq:objFunDC} is composed of differences of convex functions, 
the subtrahend \EcpkiMinus can be approximated with a linear function using the first-order Taylor series expansion~at~$\mathbf{\bar{f}}~=~\left\{ \bar{f}_n \right\}$:
\begin{twocolumn}
\begin{align}
\nonumber
   \EcpkiMinus\left( f_n\right) &\leq \EcpkiMinus\left( \bar{f}_n\right) + \left. \frac{\partial  \EcpkiMinus\left( \freqi \right)}{\partial \freqi} \right|_{\freqi = \bar{f}_n} \!\!\left( \freqi - \bar{f}_n \right)
    \\ 
    & \triangleq \EcpkiMinusTilde
    \text{.}
\end{align}
\end{twocolumn}
\begin{onecolumn}
\begin{align}
   \EcpkiMinus\left( f_n\right) &\leq \EcpkiMinus\left( \bar{f}_n\right) + \left. \frac{\partial  \EcpkiMinus\left( \freqi \right)}{\partial \freqi} \right|_{\freqi = \bar{f}_n} \!\!\left( \freqi - \bar{f}_n \right)
     \triangleq \EcpkiMinusTilde
    \text{.}
\end{align}
\end{onecolumn}
After substituting \EcpkiMinus with \EcpkiMinusTilde in \eqref{eq:objFunDC}, the objective function becomes:

\begin{twocolumn}
\begin{align}
\label{eq:objFunDCApro}
    &  \left(\mathbf{a}^\star,\mathbf{f}^\star \right) = 
    \\ \nonumber
    & \quad \arg \min_{\mathbf{a},\mathbf{f}} \sum_{r \in \R}   \sum_{n \in \F} \aki \left(\underbrace{ \EcpkiPlus - \EcpkiMinusTilde}_{\EcpkiTilde} + \Ecommki \right) \text{.}
\end{align}
\end{twocolumn}
\begin{onecolumn}
\begin{align}
\label{eq:objFunDCApro}
    &  \left(\mathbf{a}^\star,\mathbf{f}^\star \right) = 
  \arg \min_{\mathbf{a},\mathbf{f}} \sum_{r \in \R}   \sum_{n \in \F} \aki \left(\underbrace{ \EcpkiPlus - \EcpkiMinusTilde}_{\EcpkiTilde} + \Ecommki \right) \text{.}
\end{align}
\end{onecolumn}

This transformed optimization problem is convex for fixed allocation variables, thus it can be solved by employing primal and dual decomposition methods \cite{boyd,Primal}.
The primal decomposition can be applied when the problem has a coupling variable such that, when fixed to some value, the rest of the optimization problem decouples into several subproblems.
Thus, let us decompose our objective problem to:
\begin{align}
\label{eq:objTwpStep}
  \left(\mathbf{a}^\star,\mathbf{f}^\star \right) = 
    %\\ \nonumber
      \arg \min_{\mathbf{a}} \arg \min_{\mathbf{f}} \sum_{r \in \R} \sum_{n \in \F} \aki \left(\EcpkiTilde + \Ecommki \right) 
\end{align}
subject to \eqref{eq:c1}~--  { \eqref{eq:akiIn01}}.
Now, according to \eqref{eq:objTwpStep}, the solution to the optimization problem comes down to solving a two-step minimization problem.
In the first step, the optimal \ac{CPU} frequencies $\mathbf{f}^\star$ are determined for fixed allocation variables.
First, let us define the auxiliary variables $\freqi^r$ determining the \ac{CPU} frequencies of node $n$  where request $r$ is allocated. The relation between $\freqi^r$ and $\freqi$ is given by: $\freqi = \sum_{r \in \R} \aki \freqi^r$ while satisfying constraints \eqref{eq:c1} and \eqref{eq:c2}.
The optimal values of allocation variables $\mathbf{a}^\star$ are obtained in the second step based on the previously determined $\mathbf{f}^\star$ and constraints \eqref{eq:c1}~-- \eqref{eq:c2}.
Thus, we can now define the Lagrangian function of the subproblem for determining $\mathbf{f}^\star$:
\begin{onecolumn}
\begin{align}
\label{eq:lagrange}
\nonumber
    &\mathcal{L} \left( \mathbf{a},\mathbf{f}, \boldsymbol{\mu}, \boldsymbol{\Phi}, \boldsymbol{\Psi} \right) =  \sum_{r \in \R}   \sum_{n \in \F}  \aki \left(\EcpkiTilde + \Ecommki \right) + 
    \\ 
    & \qquad -  \sum_{n \in \F}   \Phi_{n} \left(\freqimin - \freqi^r \right) -  \sum_{n \in \F} \Psi_{n} \left(\freqi^r - \freqimax \right)
    -   \sum_{r \in \R} \mu^r  \left( \Dtotk - \Dmaxk\right) \text{,}
\end{align}
\end{onecolumn}
\begin{twocolumn}
\begin{align}
\label{eq:lagrange}
\nonumber
    &\mathcal{L} \left( \mathbf{a},\mathbf{f}, \boldsymbol{\mu}, \boldsymbol{\Phi}, \boldsymbol{\Psi} \right) =  \sum_{r \in \R}   \sum_{n \in \F}  \aki \left(\EcpkiTilde + \Ecommki \right) 
    \\ \nonumber
    & \qquad -  \sum_{n \in \F}   \Phi_{n} \left(\freqimin - \freqi^r \right) -  \sum_{n \in \F} \Psi_{n} \left(\freqi^r - \freqimax \right)
    \\ 
    & \qquad -   \sum_{r \in \R} \mu^r  \left( \Dtotk - \Dmaxk\right) \text{,}
\end{align}
\end{twocolumn}
and the Lagrange dual problem:
\begin{twocolumn}
\begin{align}
\label{eq:dual}
&\left(\mathbf{a}^\star,\mathbf{f}^\star, \boldsymbol{\mu}^\star,  \boldsymbol{\Phi}^\star , \boldsymbol{\Psi}^\star \right) =
\\ \nonumber
     &\qquad \qquad\arg \min_{{\boldsymbol{\mu},  \boldsymbol{\Phi} , \boldsymbol{\Psi} \geq 0}} \arg \min_{\mathbf{a}} \arg \min_{\mathbf{f}}  \mathcal{L} \left(\mathbf{a}, \mathbf{f}, \boldsymbol{\mu}, \boldsymbol{\Phi}, \boldsymbol{\Psi} \right)
       %\\ \nonumber
     %&  \text{subject to \eqref{eq:c1}, \eqref{eq:c2},} 
\end{align}
\end{twocolumn}
\begin{onecolumn}
\begin{align}
\label{eq:dual}
&\left(\mathbf{a}^\star,\mathbf{f}^\star, \boldsymbol{\mu}^\star,  \boldsymbol{\Phi}^\star , \boldsymbol{\Psi}^\star \right) =
\arg \min_{{\boldsymbol{\mu},  \boldsymbol{\Phi} , \boldsymbol{\Psi} \geq 0}} \arg \min_{\mathbf{a}} \arg \min_{\mathbf{f}}  \mathcal{L} \left(\mathbf{a}, \mathbf{f}, \boldsymbol{\mu}, \boldsymbol{\Phi}, \boldsymbol{\Psi} \right)
     %  \\ \nonumber
    % &  \text{subject to \eqref{eq:c1}, \eqref{eq:c2},} 
\end{align}
\end{onecolumn}
subject to \eqref{eq:c1}, \eqref{eq:c2}, where $\boldsymbol{\mu} = \left\{  \mu^r \right\}$, $\forall{r} \in \R$, $\mu^r \in \RealPlus$, $\boldsymbol{\Phi} = \left\{   \Phi_{n} \right\}$, $\forall{n} \in \F$, $\Phi_{n} \in \RealPlus$ and  $\boldsymbol{\Psi} = \left\{   \Psi_{n} \right\}$, $\forall{n} \in \F$, $\Psi_{n} \in \RealPlus$ are the Lagrangian multipliers responsible for fulfilling constraints \eqref{eq:delayConstraint} { and} \eqref{eq:fminfmax}, respectively.
The dual problem in \eqref{eq:dual} can be decomposed into a master problem and subproblems, and thus solved in an iterative manner.
The allocation variables $\mathbf{a}$ and \ac{CPU} frequencies $\mathbf{f}$ are obtained by solving subproblems and then the Lagrange multipliers $\boldsymbol{\mu}, \boldsymbol{\Phi}, \boldsymbol{\Psi} $ are updated by solving the master problem for the obtained frequencies.
This process continues until convergence while satisfying constraints.

\vspace{-0.5cm}
\subsection{Solving the Subproblems}
\label{sec:subproblems}
The primal problem is solved in two steps.
First, the optimal values of the \ac{CPU} frequencies $\mathbf{f}^\star$ for each request $r \in \R$ and node $n \in \F$ are obtained.
Then, in the second step, the optimal values of the allocation variables $\mathbf{a}^\star$ are determined based on $\mathbf{f}^\star$.
Thus, with \ac{KKT} conditions, for fixed allocation variables~$\mathbf{a}$, we can find the optimal \ac{CPU} frequencies  by taking the partial derivative of \eqref{eq:lagrange} with respect to $\freqi^r$ setting the gradient to 0:
\begin{align}
    \label{eq:calcFreq}
   \frac{ \partial \mathcal{L}}{ \partial {{\freqi^r}} } = 0 \qquad \forall{n \in \mathcal{F}}\text{, }  \forall{r \in \R} \text{.}
\end{align}
Due to the polynomial form of the objective function and constraint \eqref{eq:delayConstraint}, there is no closed-form solution for the above equation.
Therefore, the numerical method, \eg the Newton method with the maximum number of iterations $I_{num}$ has to be applied to solve it.   

Vector $\mathbf{a}^\star$ can be obtained based on the optimal values of the \ac{CPU} clock frequency determined in the first step by solving the following optimization problem.
%, \ie to find:
\begin{twocolumn}
\begin{align}
\label{eq:akiProblem}
    \mathbf{a}^\star=
    &\arg \max_{\mathbf{a}} \sum_{r \in \R}   \sum_{n \in \F}  \aki  \left(\EcpkiOptTilde + \Ecommki \right) 
    \\ \nonumber 
    & \qquad -  \sum_{n \in \F}   \Phi_{n} \left(\freqimin - \freqi^{r^\star} \right) -  \sum_{n \in \F} \Psi_{n} \left(\freqi^{r^\star} - \freqimax \right)
    \\ \nonumber
    & \qquad -   \sum_{r \in \R} \mu^r  \left( \Dtotk^\star - \Dmaxk\right) \text{,} %\\ \nonumber
     %&  \text{subject to \eqref{eq:c1}, \eqref{eq:c2},} 
\end{align}
\end{twocolumn}
\begin{onecolumn}
\begin{align}
\label{eq:akiProblem}
    \nonumber
    \mathbf{a}^\star=
    &\arg \max_{\mathbf{a}} \sum_{r \in \R}   \sum_{n \in \F}  \aki  \left(\EcpkiOptTilde + \Ecommki \right) 
    \\ 
    & \qquad -  \sum_{n \in \F}   \Phi_{n} \left(\freqimin - \freqi^{r^\star} \right) -  \sum_{n \in \F} \Psi_{n} \left(\freqi^{r^\star} - \freqimax \right)
    -   \sum_{r \in \R} \mu^r  \left( \Dtotk^\star - \Dmaxk\right) \text{,} %\\ \nonumber
     %&  \text{subject to \eqref{eq:c1}, \eqref{eq:c2},} 
\end{align}
\end{onecolumn}
{ subject to \eqref{eq:c1}, \eqref{eq:c2}}, where $\EcpkiOptTilde = \EcpkiTilde\left( \freqi^{r^\star}\right) $ and $ \Dtotk^\star =\Dtotk\left( \freqi^{r^\star}\right)$.
The~optimization problem defined in \eqref{eq:akiProblem} is the linear assignment problem, and can be solved by the Hungarian algorithm \cite{Hungarian}.
Let us define matrix $\mathbf{\Theta} = \left\{ \EtotkiOpt \right\}$, $\forall{r} \in \R$ and $\forall{n} \in \mathcal{C}$ with $\left| \R \right|$ rows and $\left| \mathcal{C} \right|$ columns, and matrix $\mathbf{\Lambda} = \left\{ \EtotkiOptTilde \right\}$, $\forall{r} \in \R$ and $\forall{n} \in \mathcal{F}$ with the same number of rows and $\left| \mathcal{F} \right|$ columns, where $\EtotkiOptTilde = \EcpkiOptTilde + \Ecommki$.
To reflect unlimited computational resources at each \ac{CN}, we introduce matrix $\mathbf{\Omega} = \left[  \mathbf{\Lambda} \:\mathbf{\Theta} \otimes \mathbf{1}_{1 \times \left| \R \right|} \right]$, 
where $\otimes$ is the Kronecker tensor product while $\mathbf{1}_{1 \times \left| \R \right|}$ is a vector of ones with one row and~$ \left| \R \right|$ columns.
It means that the columns of { $\mathbf{\Theta}$} are replicated~$ \left| \R \right|$ times and matrix $\mathbf{\Omega}$ has $\left| \R \right|$ rows and $|\R|\cdot|\C|+|\F|$ columns.
\Eg, in the case of three tasks $\left| \R \right| = 3$, two fog nodes $\F=\{1,2\}$ and one cloud $\C=\{3\}$, the matrix $\boldsymbol{\Omega}$ is defined as follows:
\begin{align}
\label{eq:HungarianOmegaMatrixExample}
\boldsymbol{\Omega} = 
\begin{bmatrix} 
{\tilde{E}{_{\text{tot},1}^{1}}}^\star 
 & 
{\tilde{E}{_{\text{tot},2}^{1}}}^\star
&
{E{_{\text{tot},3}^{1}}}^\star 
 & 
{E{_{\text{tot},3}^{1}}}^\star
& 
{E{_{\text{tot},3}^{1}}}^\star
\\ 
{\tilde{E}{_{\text{tot},1}^{2}}}^\star  
& 
{\tilde{E}{_{\text{tot},2}^{2}}}^\star
&
{E{_{\text{tot},3}^{2}}}^\star  
& 
{E{_{\text{tot},3}^{2}}}^\star
& 
{E{_{\text{tot},3}^{2}}}^\star
\\ 
\undermat{\EtotkiOpt \: \forall{n \in \F} }{{\tilde{E}{_{\text{tot},1}^{3}}}^\star  
& 
{\tilde{E}{_{\text{tot},2}^{3}}}^\star}
&
\undermat{  \EtotkiOpt \: \forall{n \in \mathcal{C}}}{{E{_{\text{tot},3}^{3}}}^\star  
& 
{E{_{\text{tot},3}^{3}}}^\star
& 
{E{_{\text{tot},3}^{3}}}^\star}
\end{bmatrix}
\\ \nonumber
\end{align}
Next, applying the Hungarian algorithm for matrix $\mathbf{\Omega}$ the matrix with binary values is determined, \eg:
\begin{align}
\label{eq:omegaMatrix}
\mathcal{H}\left(\boldsymbol{\Omega}\right) = 
\begin{bmatrix} 
0  & 1 & 0 & 0 & 0
\\ 
0  & 0 & 1 & 0 & 0
\\ 
 0 & 0 & 0 & 0 & 1
\end{bmatrix}
%\\ \nonumber
\end{align}
The example above shows that the first task is computed in the second fog node while the second and the third task are computed in the \ac{CN}.
Thus, the optimal values of $a{_n^r}^\star$ can be determined by: 
\begin{equation}
a_n^{r^\star} =
\left\{ 
\begin{array}{lcl}
\mathcal{H}\left(\boldsymbol{\Omega}\left(r,n\right)\right) & \mathrm{for} &   n \leq \left|F \right| \\
\sum_{j = \left|F \right| +1}^{|\F|+|\R||\C| + 1}  \mathcal{H}\left( \boldsymbol{\Omega} \left(r,j\right) \right) & \mathrm{for} &  n > \left|F \right| 
\end{array}
\right. \text{.} \label{eq:akiOpt}
\end{equation}
\vspace{-0.5cm}
\subsection{Solving the Master Problem}
\label{sec:master_problem}
We can fulfill constraints \eqref{eq:delayConstraint} and \eqref{eq:fminfmax} by determining the search range of the optimal solution.
Let \freqidelay denote the minimum value of \freqi which satisfies the constraint \eqref{eq:delayConstraint} for request $r \in \R$ processed at node $n \in \F$.
These can be obtained by solving the following equation:
\begin{align}
    \label{eq:Dtotki}
    \Dtotki - \Dmaxk = 0, \quad \forall{n \in \F}, \: \forall{r \in \R} \text{.}
\end{align} 
Inserting \Dtotki from \refEq{eq:Dtotk} (together with \Dcpki taken from \refEq{eq:Dcpk}) into \refEq{eq:Dtotki} we get:

\begin{twocolumn}
\begin{align}
\label{eq:minDelayFreq}
    \freqi^r &= \frac{\Lk \thetak}{\si \left(\Dmaxk - \Dcommki - \Dqueueki \right)}
    \\ \nonumber
    &\triangleq \freqidelay
    \text{.}
\end{align}
\end{twocolumn}
\begin{onecolumn}
\begin{align}
\label{eq:minDelayFreq}
   \freqidelay &= \frac{\Lk \thetak}{\si \left(\Dmaxk - \Dcommki - \Dqueueki \right)}
    \text{.}
\end{align}
\end{onecolumn}
If $\freqidelay > f_{\mathrm{max},n}$, the task $r$ cannot be processed in \ac{FN} $n$ within the delay constraint.
If $\freqidelay  \leq f_{\mathrm{min},n}$, the minimum \ac{CPU} clock frequency is kept at $f_{\mathrm{min},n}$. %set to $\freqidelay =f_{\mathrm{min},n}$. 
Thus, if the obtained optimal clock frequency is in the range $\freqi \in \langle \max \left\{ f_{\mathrm{min},n}, \freqidelay \right\}, f_{\mathrm{max},n} \rangle$,
the Lagrange { multipliers} in \eqref{eq:lagrange} and \eqref{eq:akiProblem} simplify by setting $\Phi_n = 0 $, $\Psi_n = 0 $, $\forall{n \in \F}$ and $\mu^r = 0 $, $\forall{r \in \R}$. 

Finally, we propose the algorithm called \ac{EEFFRA} (\refAlg{alg:ee-resource-allocation}) for finding the solution (\ac{CPU} clock frequencies and request allocation over the nodes) for total energy minimization with delay constraints, as discussed above.
The computational complexity of the proposed
algorithm results from the complexity of the Hungarian algorithm (line \ref{alg:hungarian}) and the iterative frequency finding procedure (lines \ref{alg:repeat}-\ref{alg:until}).
The complexity of the Hungarian algorithm is proportional to a cube of greater number: number of tasks or number of agents.
In our model, the requests (tasks) can be assigned to the fog nodes or the cloud %(\REM{which represent the} agents).
(agents).
Moreover, the cloud can process more than one request  simultaneously.
Therefore, in our model, we have $|\R|$ tasks which can be assigned to the $|\F|+|\R||\C|$ nodes, where $|\R||\C|$ represents the cloud nodes which can process more than one request.

Thus, in our solution the complexity of the Hungarian algorithm equals 
$\mathcal{O}\left( \left( |\F|+|\R||\C|\right) ^3\right)$.
The second part of the computational complexity of the proposed \ac{EEFFRA} algorithm results from the \ac{CPU} frequency calculation.
Let us observe that the \ac{CPU} frequency has to be calculated for each node and each request \ie $ \left|\R\right|  \left|\mathcal{N}\right|$ frequencies have to be determined.
The main step of the proposed algorithm (line \ref{alg:numer}) determines the optimal \ac{CPU} frequencies for a given approximation of the objective function which are then updated in the loop (line \ref{alg:frequpdate}).
This procedure is repeated until the termination conditions are met (line \ref{alg:until}).
Thus, the complexity of the \ac{CPU} frequency calculation is equal to $\mathcal{O}\left(  \left|\R\right|  \left|\mathcal{N}\right|i_\mathrm{num} i_\mathrm{sca}\right)$, where $i_\mathrm{num}$ and $i_\mathrm{sca}$ are the numbers of iterations of the numerical method applied to solve \eqref{eq:calcFreq} and the \ac{SCA} method, respectively.
Complexity of the entire EEFFRA algorithm is therefore equal to $\mathcal{O}\left( \left( |\F|+|\R||\C|\right)^3+  \left|\R\right|  \left|\mathcal{N}\right|i_\mathrm{num} i_\mathrm{sca}\right) $.

% ================================
	%\renewcommand{\arraystretch}{0.7}
\begin{algorithm}
\caption{The \ac{EEFFRA} in the fog computing networks.}

\label{alg:ee-resource-allocation}
\begin{algorithmic}[1]
\linespread{1.35}\selectfont

\STATE \textbf{Inputs:} \Lk, \thetak, \ok, \Dmaxk for $r \in \R$, $\{ p_{n,0},p_{n,3},...,p_{n,Q}\}$, \freqimin, \freqimax, \si, \di for $n \in \F$, \gammaik, \ribk for $r \in \R$ and $n \in \F$ and \rback, \chiRTT,  maximum number of iterations $I_{\mathrm{num}}$, $I_{\mathrm{sca}}$, iteration index 
$i_\mathrm{sca}$, maximum error $\varepsilon$ and initial values of optimization variables $\mathbf{\bar{f}}$
\STATE \textbf{Outputs:} \Etotki for $r \in \R$ and $n \in \F$, $\mathbf{f}^\star$, $\mathbf{a}^\star$ 
\REPEAT
\label{alg:repeat}

%\REPEAT
\STATE calculate $\mathbf{f}^\star$ by solving \eqref{eq:calcFreq}
in the range $\freqi~\in~\langle \max \left\{ f_{\mathrm{min},n}, \freqidelay \right\}, f_{\mathrm{max},n} \rangle  $ for $\Phi_n = 0 $, $\Psi_n = 0 $, $\forall{n \in \F}$ and $\mu^r = 0 $, $\forall{r \in \R}$ using 
numerical method with max. $I_\mathrm{num}$ iterations%
\label{alg:numer}

\STATE $  \mathbf{\bar{f}} \leftarrow \mathbf{f}^\star$
\label{alg:frequpdate}
\STATE $i_\mathrm{sca}\leftarrow i_\mathrm{sca} + 1$ 
\UNTIL{$\left| \mathbf{\bar{f}} - \mathbf{\bar{f}}  \right| \leq \varepsilon$ \OR $i_\mathrm{sca} = I_\mathrm{sca}$}
\label{alg:until}

\STATE calculate $\mathbf{a}^\star$ using the Hungarian method for the matrix $\mathbf{\Omega}$ and \eqref{eq:akiOpt}
\label{alg:hungarian}

\STATE $\Etotki \leftarrow \EtotkiTilde$ for $r \in \R$ and $n \in \F$

\end{algorithmic}

\end{algorithm}   
% ================================
\vspace{-0.9cm}
\subsection{Low-complexity solution (LC-\ac{EEFFRA})}
\label{sec:low-compl-sol}
In the following approach to our optimization problem called Low-Complexity \ac{EEFFRA} (LC-\ac{EEFFRA}), we remove the Hungarian algorithm from \ac{EEFFRA} leading to reduced computational complexity to $\mathcal{O}\left(\left|\R\right|  \left|\mathcal{N}\right| i_\mathrm{num} i_\mathrm{sca}\right) $.
The optimal values of frequency $\mathbf{f^\star}$ are determined in the same way as in \refAlg{alg:ee-resource-allocation} while the values of $\mathbf{a^\star}$ are obtained in a heuristic manner.
In this heuristic approach, only a single request $r \in \R$ is considered at a time.
It is allocated to node ${n}^\star$ where the energy consumption for processing $r$ is the lowest, \ie we find:
\begin{align}
{n}^\star = \arg\min_{n} \EtotkiOpt \: \forall{r \in  \R } \text{.}
\end{align}
Values \ti are updated after allocation of each request to prevent multiple collisions of two or more requests at the same \ac{FN}.
The examination order of requests arriving at the same time is random to emulate a decentralized approach.

% !TEX root = main.tex

\section{Results}
\label{sec:results}

Results obtained from computer simulations are presented in this section.
Let us consider a network with $|\F|=10$ \acp{FN} and $|\C|=1$ cloud \ac{DC}.
Simulation parameters are summarized in \refTab{tab:ranges}.
The process of generating requests for simulations is as follows.
At time $\tk \in \T$ there appear between 5 and 10 (uniform distribution) new computational requests. 
The value $\tk$ is generated at a random delay after previous time instance $T_{k-1}$.
The difference $\tk - T_{k-1}$ is chosen to be a random variable of exponential distribution with average value 50~ms (intensity = 20~$\text{s}^{-1}$).
The requests have randomly (with uniform distribution) assigned values of their parameters (size, arithmetic intensity, delay requirement) in ranges shown in \refTab{tab:ranges}.

\begin{figure}
	\centering
	\includegraphics[width=0.4\textwidth]{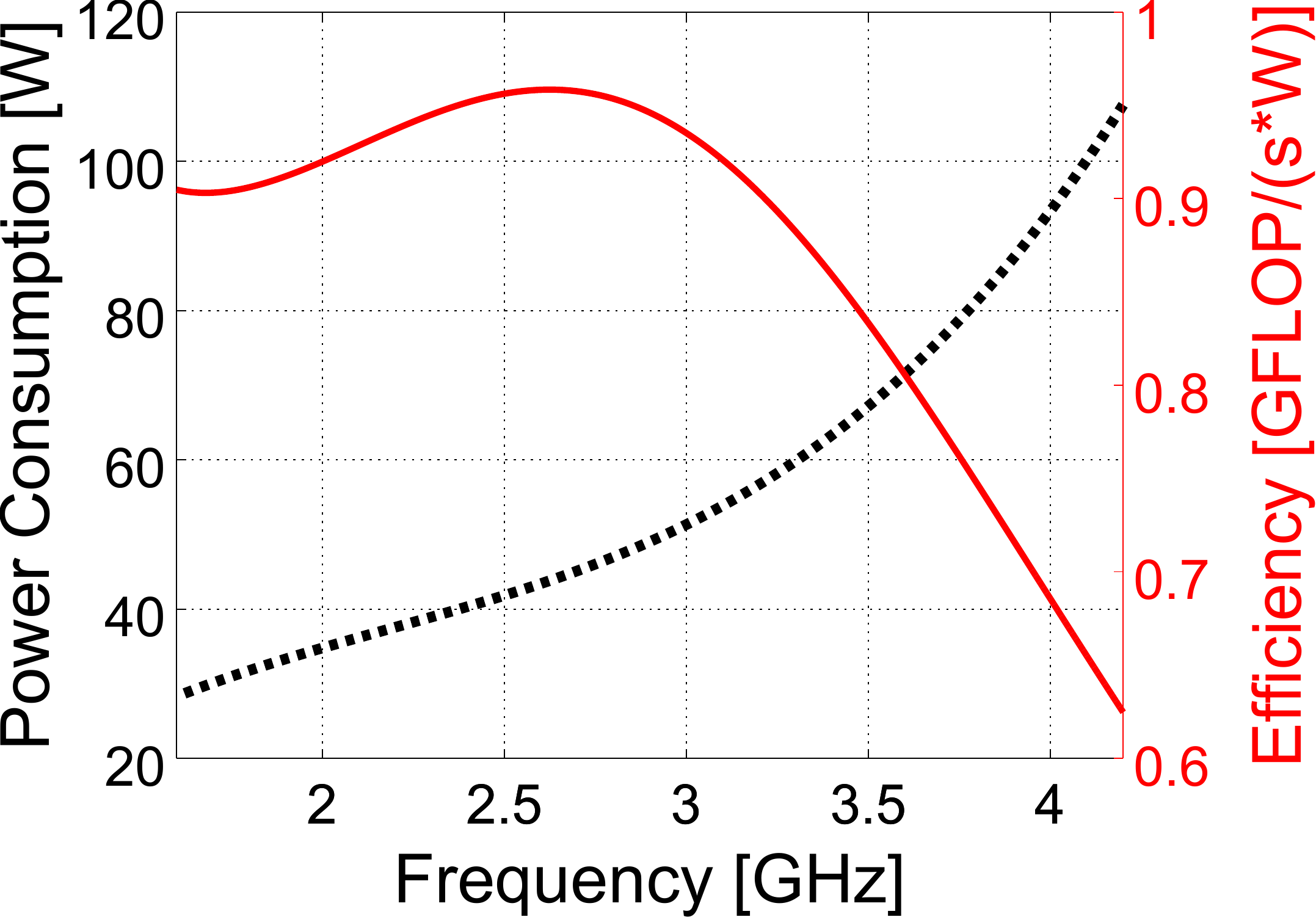}	
	\caption{Power consumption and energy efficiency of Intel Core i5-2500K vs. CPU frequency.}
		\label{fig:beta_power_new}
			\vspace{-1cm}	

\end{figure}

It is assumed that each \ac{FN} uses a single Intel Core i5-2500K as its \ac{CPU}.
Data relating frequency, voltage, and power consumption of i5-2500K is taken from \cite{Wong2012} and fit into \refEq{eq:beta3} adopted from \cite{Park2013}. The resulting power consumption and energy efficiency are plotted in \refFig{fig:beta_power_new}.
These figures show that power consumption increases faster-than-linearly with operating frequency and that the frequency with the highest energy efficiency is around 2.6~GHz.
The cloud \acp{CPU} are parameterized according to the \textit{Intel Xeon Phi} family commonly used in computer clusters \cite{Intel,Green2020} characterized with $s=32\text{ FLOP/cycle}$ \cite{Dolbeau2018}, and run at a constant frequency $1.5$~GHz.
Transmission parameters are analogous to those used in \cite{Kopras2019}. 
If not stated differently, simulations for each data point are obtained over 550 time instances $T_k$.
Results from the first 50 instances are discarded.
Random number generator seeds are kept the same for each value of swept parameters for a fair comparison of results.  

Our solutions, \ie \ac{EEFFRA} and LC-\ac{EEFFRA}, are compared with three reference methods.
The first method, called \textit{Cloud Only}, processes all requests in the cloud.
The second method, called \textit{Fog Only}, processes all requests in the \acp{FN}.
The \acp{FN}' \ac{CPU} frequencies and requests-to-nodes assignments are determined using LC-\ac{EEFFRA}.
Finally, the third method, called \textit{Fog Simple}, processes  requests in the same \ac{FN} that these requests arrived at.
Still, it uses optimal \acp{FN}' \ac{CPU} frequencies determined using LC-\ac{EEFFRA}.   
Simulations are performed using MATLAB.% and Octave.

\begin{twocolumn}
\begin{table}
	\caption{Simulation parameters.}
	\label{tab:ranges}
	%\vspace{-2mm}
	\centering
	\begin{tabular}{p{1.3cm} p{2.2cm} |p{1.3cm} p{2.2cm}}
		\hline
		 Symbol & Value/Range & Symbol & Value/Range\\
		\hline
		\multicolumn{4}{c}{\textbf{Requests}, $r \in \R$}\\
		\hline
	 \Lk %& size of request %$r \in \R$ 
	 &[1,10] MB&
		 \thetak %& arithmetical intensity of task %$r \in \R$ 
		 &[1,100] FLOP/bit\\
		 
		 \ok %& output-to-input data size ratio of request %$r \in \R$
		 & [0, 0.5] &
	 \Dmaxk %& maximum tolerated delay requirement for request %$r \in \R$
	 & [100, 1000] ms\\
	 $\lvert \R \rvert$ & [5,10] \\
	 \hline
	 	\multicolumn{4}{c}{\textbf{Number of nodes}}\\
		\hline
	
	$\lvert \F \rvert$ & 10&
		 $\lvert \C \rvert$ & 1\\
	
	 \hline
	 		\multicolumn{4}{c}{\textbf{Computations in the fog} \cite{Wong2012,Dolbeau2018,Park2013}, $n\in\F$ }\\
	 		\hline
	 			 \powerParamThreeI, \powerParamTwoI % & %parameters of \Pacti polynomial 
	 			 & 5.222, 34.256 &
	 \powerParamOneI, \powerParamZeroI %&  
	 & 88.594, -47.152\\
	 \freqimin %& minimum frequency of CPU 
	 & 1.6~GHz&
	 \freqimax %& maximum frequency of CPU 
	 & 4.2~GHz\\
	 \si %& %number of FLOPs performed within a single clock cycle 
	 %number of FLOPs performed within a single clock cycle 
	 & 16 FLOP/cycle\\
	 \hline
%	 $\max(\betai)$ & maximum performance per WwWwWwatt & 0.96~GFLOP/(s$\cdot$W) @2.606~GHz \\
%\hline
	 \multicolumn{4}{c}{\textbf{Computations in the cloud} \cite{Green2020,Dolbeau2018}, $n\in\C$}\\
	 \hline
	 \freqi %& frequency of CPU 
	 & 1.5~GHz &
	% \freqimin & minimum frequency of CPU & 2.1~GHz\\
%	 \freqimax & maximum frequency of CPU & 2.1~GHz\\
	 \si %& number of FLOPs performed within a single clock cycle 
	 & 32 FLOP/cycle\\
%	\betai %& performance per watt 
%	& 1.3 %[0.5,~5.0] 
%	GFLOP/(s$\cdot$W) \\
	\hline
	 	 	\multicolumn{4}{c}{\textbf{Transmission} \cite{Olbrich2009,Bertoldi2017,VanHeddeghem2012}}\\ \hline
	 \di, $n\in\C$%&  fiberline distance to cloud 
	 & 2000~km&
	 	 \chiRTT %& delay/distance parameter
	 & 7.5~\textmu s/km \\
	 \rback %& transmission rate in the backhaul and backbone network 
	 & 1~Gbps&
		 \ribk, $n\in\F$ %&  transmission rate between fog Nodes 
		 & 1~Gbps \\

	 \gammaik, $n\in\F$ %& cost of transmitting data between FNs 
	 & 0.3 nJ/(bit$\cdot$hop) &
	 \gammaik, $n\in\C$ %& cost of transmitting data to cloud 
	 & 10 nJ/bit \\

	\hline
	\end{tabular}
\end{table}
\end{twocolumn}
\begin{onecolumn}
\begin{table}
	\renewcommand{\arraystretch}{0.7}

	\caption{Simulation parameters.}
	\label{tab:ranges}
	\vspace{-4mm}
	\centering
	\begin{tabular}{p{1.5cm} p{2.2cm} |p{1.5cm} p{2.2cm}|p{1.5cm} p{2.2cm}}
		\hline
		 Symbol & Value/Range & Symbol & Value/Range & Symbol & Value/Range\\
		\hline
		\multicolumn{6}{c}{\textbf{Requests}, $r \in \R$}\\
		\hline
	 \Lk %& size of request %$r \in \R$ 
	 &[1,10] MB&
		 \thetak %& arithmetic intensity of task %$r \in \R$ 
		 &[1,100] FLOP/bit&
		 
		 \ok %& output-to-input data size ratio of request %$r \in \R$
		 & [0, 0.5] \\
	 \Dmaxk %& maximum tolerated delay requirement for request %$r \in \R$
	 & [100, 1000] ms&
	 	 $\lvert \R \rvert$ & [5,10] &$\overline{\tk - T_{k-1}}$ & 50~ms\\
	 \hline
	 	\multicolumn{6}{c}{\textbf{Number of nodes}}\\
		\hline
	
	$\lvert \F \rvert$ & 10&
		 $\lvert \C \rvert$ & 1\\
		\hline 
	 		\multicolumn{6}{c}{\textbf{Computations in the fog} \cite{Wong2012,Dolbeau2018,Park2013}, $n\in\F$ }\\
	 		\hline
	    \si %& %number of FLOPs performed within a single clock cycle 
	    %number of FLOPs performed within a single clock cycle 
	    & 16 FLOP/cycle &
	 \powerParamThreeI, \powerParamTwoI % & %parameters of \Pacti polynomial 
	 & 5.222, 34.256 &
	 \freqimin %& minimum frequency of CPU 
	 & 1.6~GHz \\
	 $Q$ & 3 &
	 \powerParamOneI, \powerParamZeroI %&  
	 & 88.594, -47.152 &
	 \freqimax %& maximum frequency of CPU 
	 & 4.2~GHz\\
	 \hline
%	 $\max(\betai)$ & maximum performance per watt & 0.96~GFLOP/(s$\cdot$W) @2.606~GHz \\
%\hline
	 \multicolumn{6}{c}{\textbf{Computations in the cloud} \cite{Green2020,Dolbeau2018}, $n\in\C$}\\
	 \hline
	 \freqi %& frequency of CPU 
	 & 1.5~GHz &
	% \freqimin & minimum frequency of CPU & 2.1~GHz\\
%	 \freqimax & maximum frequency of CPU & 2.1~GHz\\
	 \si %& number of FLOPs performed within a single clock cycle 
	 & 32 FLOP/cycle\\
%	\betai %& performance per watt 
%	& 1.3 %[0.5,~5.0] 
%	GFLOP/(s$\cdot$W) \\
	\hline
	 	 	\multicolumn{6}{c}{\textbf{Transmission} \cite{Olbrich2009,Bertoldi2017,VanHeddeghem2012}}\\ \hline
	 \di, $n\in\C$%&  fiberline distance to cloud 
	 & 2000~km&
	 	 \chiRTT %& delay/distance parameter
	 & 7.5~\textmu s/km &
	 \rback %& transmission rate in the backhaul and backbone network 
	 & 1~Gbps \\
		 \ribk, $n\in\F$ %&  transmission rate between fog Nodes 
		 & 1~Gbps &

	 \gammaik, $n\in\F$ %& cost of transmitting data between FNs 
	 & 0.3 nJ/(bit$\cdot$hop) &
	 \gammaik, $n\in\C$ %& cost of transmitting data to cloud 
	 & 10 nJ/bit \\

	\hline
	\end{tabular}
	\vspace{-1cm}
\end{table}

\end{onecolumn}
%\vspace{-2cm}
% ========================================

\vspace{-0.5cm}
\subsection{Convergence of Algorithms and Optimality of Solution}
\ac{EEFFRA} utilizes \ac{SCA} for finding optimal operating frequencies and the Hungarian algorithm for assigning requests to nodes. 
The Hungarian algorithm is guaranteed to find the optimum in polynomial time \cite{Hungarian}. 
\ac{SCA} is guaranteed to converge \cite{Zappone2017}.
To show the \ac{SCA} convergence rate, we plot normalized energy costs resulting from offloading requests depending on the maximum number of algorithm iterations in \refFig{fig:convergence}.
Apart from \ac{SCA} iterations $I_\mathrm{sca}$, we also vary the maximum number of iterations $I_\mathrm{num}$ used to find the optimum \ac{CPU} frequencies in \refAlg{alg:ee-resource-allocation}.
For clarity of convergence analysis, we assume that there is no cloud, \ie $\lvert \C \rvert = 0$.
The operating frequency of the cloud is not adjusted (not a variable) and therefore does not influence the analysis. 
Normalization is obtained by plotting the relative difference between the cost achieved by \ac{EEFFRA} and the optimal cost found by solving the original problem without \ac{SCA}. 
{Since the degree of polynomial modeling \ac{CPU} power consumption in these simulations equals 3, the optimal frequencies $\freqi^{r^\star}$ can be found analytically for each request $r \in \R$ and node $n \in \F$ which result in the lowest cost $\Ecpki$ while fulfilling \eqref{eq:delayConstraint}.
Since the first derivative of \Ecpki over $f_n$ from \eqref{eq:Ecp_parameters} %has at most 3 real roots for $f_n>0$, 
is continuous everywhere except at singularity at $\freqi=0$, and has at most 3 real roots,
the optimal frequency $\freqi \in \langle \max \left\{ f_{\mathrm{min},n}, \freqidelay \right\}, f_{\mathrm{max},n} \rangle$ is either obtained for the endpoint of this interval or for one of the roots of $\frac{d}{d \freqi}\Ecpki(\freqi)$.
The lowest value \Ecpki for these frequencies determines the optimal frequency $\freqi^{r^\star}$.
The solution is continued as described in \refSec{sec:solution} from \refEq{eq:akiProblem}.
It is visible in \refFig{fig:convergence} that \ac{EEFFRA} converges both quickly and to values close to the optimal ones.

\begin{figure}
	\centering
	\includegraphics[width=0.4\textwidth]{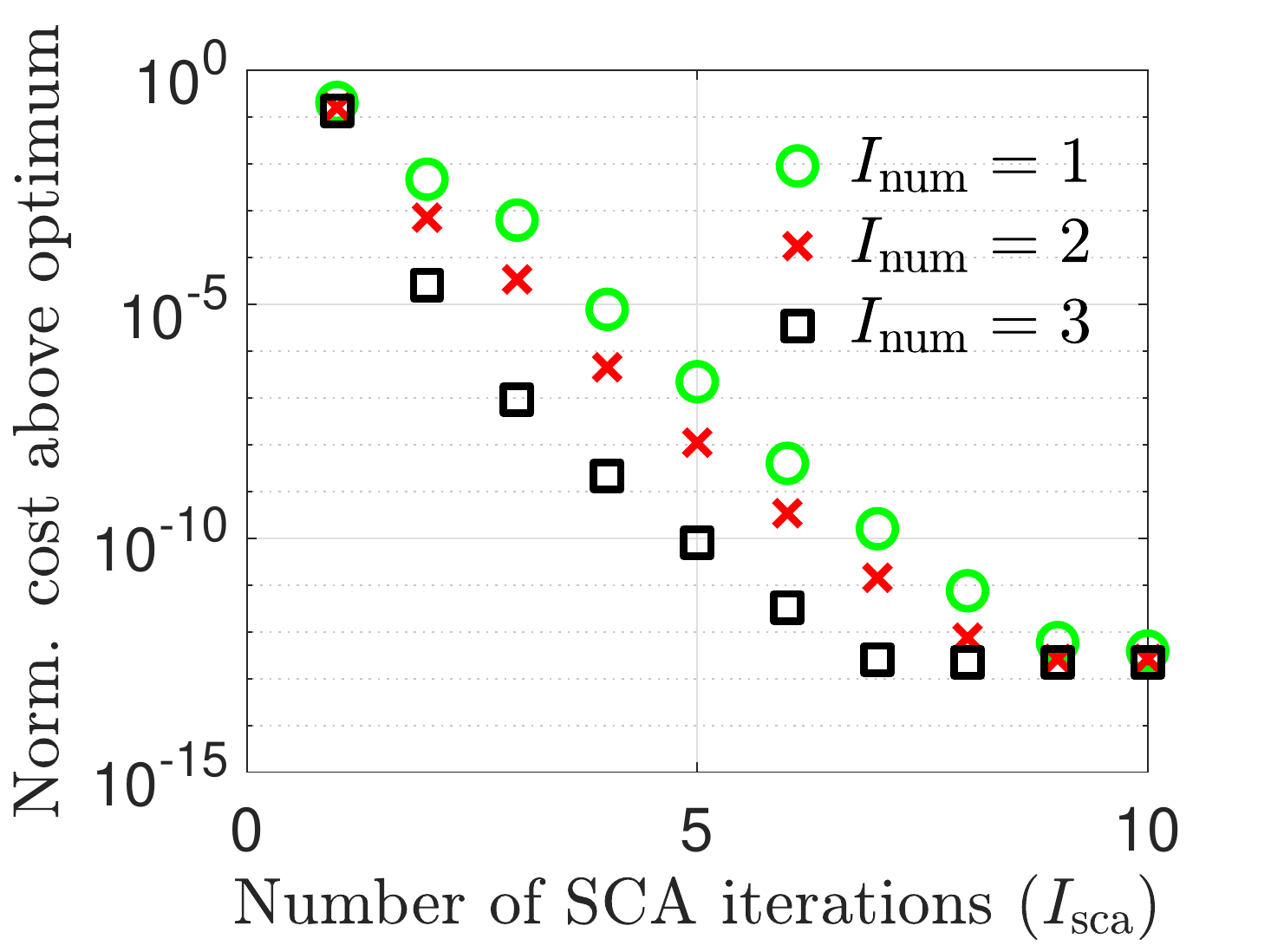}
	\caption{ Convergence of solutions found by EEFFRA to optimum with number of iterations.
	}

	\label{fig:convergence}
	\vspace{-0.9cm}
		
\end{figure}

\vspace{-0.5cm}
\subsection{Impact of Computational Energy Efficiency of the Cloud}
\label{sec:i5-1}
% ==========================================
\begin{figure}
	\centering
	\includegraphics[width=0.4\textwidth]{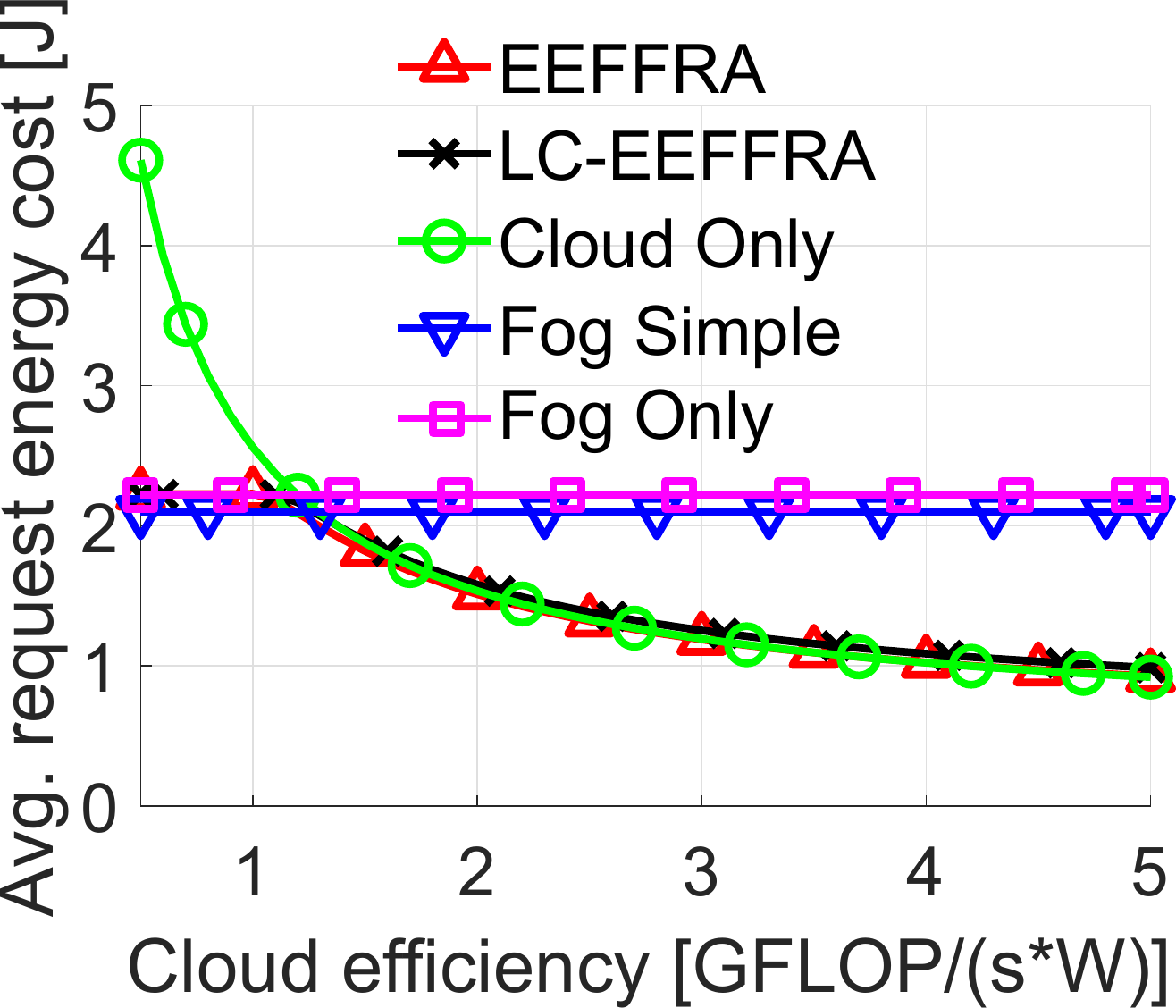}
 	\caption{Influence of cloud energy efficiency on average energy cost for chosen policies.}
	\label{fig:energy_losses}
	\vspace{-1cm}
\end{figure}
% ==========================================
First, we vary the values of the computational efficiency of the cloud \ac{DC} in the range [0.5,~5.0]~GFLOP/(s$\cdot$W) (median value for 500 of the most powerful commercially available computer clusters is 2.962~GFLOP/(s$\cdot$W) \cite{Green2020}).
The average energy costs per successfully processed request are shown in \refFig{fig:energy_losses}.
At low computational efficiency of the cloud, policies utilizing only nodes in the fog tier (\textit{Fog Simple}, \textit{Fog Only}) perform similarly to those utilizing both the fog and the cloud.
\textit{Cloud only} approach has the highest energy consumption at low efficiency of the cloud (up to around 1.3~GFLOP/(s$\cdot$W)).
Above that level \textit{Cloud only} is characterized with lower energy consumption than \textit{Fog Simple} and \textit{Fog Only} solutions and similar to  \ac{EEFFRA} and LC-\ac{EEFFRA} solutions  as under these parameters it is the most efficient to process most requests in the cloud.
\ac{EEFFRA} is slightly more efficient than LC-\ac{EEFFRA} at higher cloud efficiency values.
The percentage of requests which were unable to be processed using each of the offloading policies is the following:
The \textit{Fog Simple} solution where \acp{FN} cannot ``share'' computational requests between themselves has the highest ratio of rejected requests (8.2\%)
, while only 1.4\%-1.7\% 
requests (percentage varies depending on cloud efficiency -- lower for higher efficiency) are rejected by both proposed solutions utilizing both fog and cloud (\ac{EEFFRA} and LC-\ac{EEFFRA}).
\textit{Fog Only} and \textit{Cloud Only} have rejection rates of 1.9\% and 4.3\% respectively.
Requests which are rejected tend to have larger sizes and higher arithmetic intensities (as shown later in \refFigs{fig:hist2D131130:rejectedRequests} and \ref{fig:size}).
It ``artificially'' decreases the average-per-request cost of methods with higher rejection rates in \refFig{fig:energy_losses}, \eg causing \textit{Fog Simple} to show lower average cost than \textit{Fog Only}).

Let us examine more closely where \ac{EEFFRA} chooses to offload computational requests and what parameters impact these decisions.
\refFig{fig:hist2D131130} shows histograms of parameters characterizing offloaded requests obtained after running simulations for 2000 \tk instances.
\refFig{fig:hist2D131130}a and \refFig{fig:hist2D131130}b show  the probabilities of requests being processed in the fog tier of the network and those rejected due to too low delay requirement at cloud efficiency of 1.3~GFLOP/(s$\cdot$W).
A similar histogram for requests processed in the cloud would be superfluous as probabilities for results processed in the fog, in the cloud, and those rejected sum to 1. 
Unsurprisingly, \refFig{fig:hist2D131130}b shows that results with strict latency requirements are less likely to be successfully processed in time.
In \refFig{fig:hist2D131130}a up to 40\% of high arithmetical intensity tasks with delay requirements of around 200~ms are processed in the fog tier as a result of the cloud being unable to fulfill these requirements.
In \refFig{fig:hist2D131130}a one can see a ``threshold'' between 40 and 50 FLOP/bit below which requests are chosen by \ac{EEFFRA} to be served by the \acp{FN} rather than the cloud.
Similar histograms plotted for other values of cloud efficiencies show that this threshold increases with a less efficient cloud and decreases with a more efficient cloud.
In particular, for efficiencies below 1.0~GFLOP/(s$\cdot$W), all requests are processed in Fog. On the other hand, even for infinitely high efficiencies of the cloud around 20\% of tasks remain processed in the fog tier (low-intensity ones for which the cost of transmission to the cloud outweighs computational costs in the fog and those with strict delay requirements).
up to 40\% of high arithmetical intensity tasks with delay requirements of around 200~ms are processed in the fog tier as a result of the cloud being unable to fulfill these requirements.

% ==========================================
\begin{twocolumn}
\begin{figure}
	\centering
	\begin{subfigure}{.24\textwidth}
		\centering
			\includegraphics[width=0.999\textwidth]{plots/Hist2DinFogB13delReq.pdf}
		\caption{Histogram of requests processed in Fog Nodes (Cloud eff.: 1.3~GFLOP/(s$\cdot$W))}
	\end{subfigure}%\hfill%	\textbf{}
	\hspace{0.5mm}
	\begin{subfigure}{.24\textwidth}
		\centering
		\includegraphics[width=0.999\textwidth]{plots/Hist2DrejB13delReq.pdf}
		\caption{Histogram of rejected requests (Cloud eff.: 1.3~GFLOP/(s$\cdot$W), note scaled legend).}
	\end{subfigure}

	\centering
	\begin{subfigure}{.232\textwidth}
		\centering
			\includegraphics[width=0.999\textwidth]{plots/Hist2DinFogB11delReq.pdf}
		\caption{Histogram of requests processed in Fog Nodes (Cloud eff.: 1.1~GFLOP/(s$\cdot$W))}
	\end{subfigure} %\hfill%	\textbf{}
	\hspace{0.5mm}
	\begin{subfigure}{.232\textwidth}
		\centering
		\includegraphics[width=0.999\textwidth]{plots/Hist2DinFogB30delReq.pdf}
		\caption{Histogram of requests processed in Fog Nodes (Cloud eff.: 3.0~GFLOP/(s$\cdot$W))}
	\end{subfigure}
		\caption{Arithmetical intensity and delay requirement of requests processed in fog, and those rejected (\ac{EEFFRA}).}
	
	\label{fig:hist2D131130}
\end{figure}
\end{twocolumn}
\begin{onecolumn}
\begin{figure}[t!]
\captionsetup{justification=centering}
	\centering
	\begin{subfigure}[t]{.31\textwidth}
		\centering
			\includegraphics[width=0.999\textwidth]{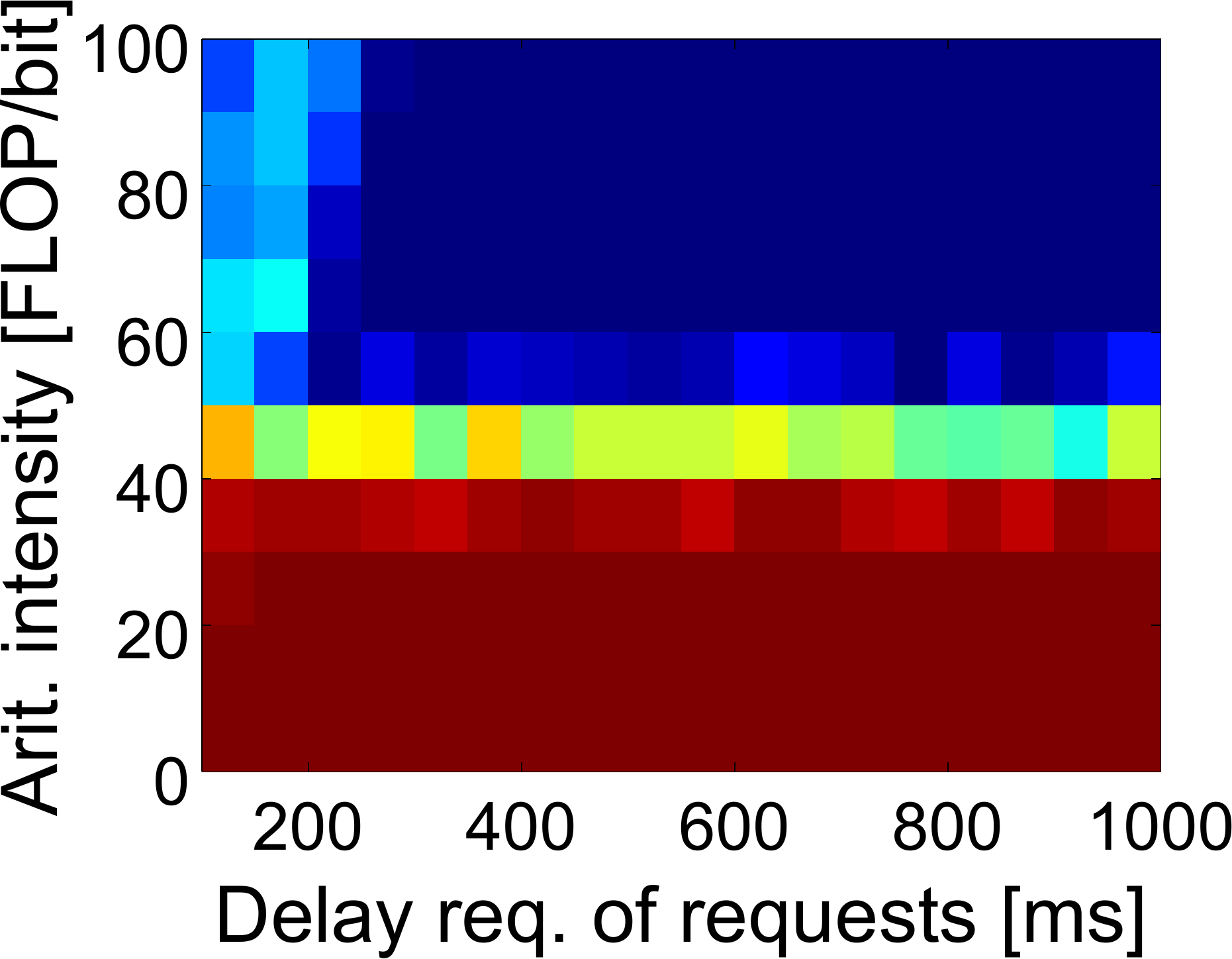}
			\caption{Requests processed in FNs}
	    \label{fig:hist2D131130:processedRequests}
	\end{subfigure}%\hfill%	\textbf{}
	\hspace{0.1mm}
	\begin{subfigure}[t]{.31\textwidth}
		\centering
		\includegraphics[width=0.999\textwidth]{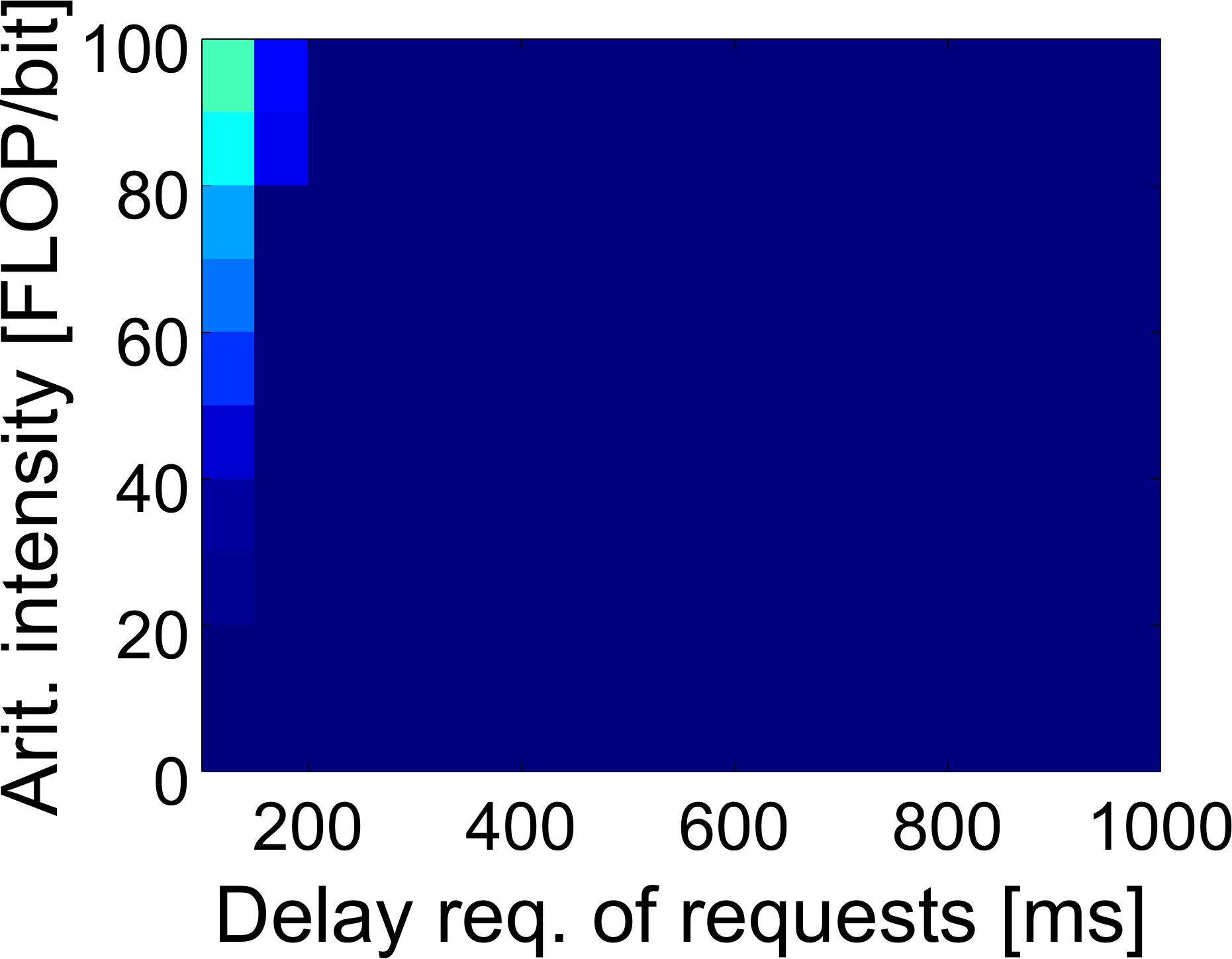}
		\caption{Rejected requests.}
	    \label{fig:hist2D131130:rejectedRequests}
	\end{subfigure}
	\begin{subfigure}[t]{.04\textwidth}
		\centering
	\includegraphics[width=0.999\textwidth]{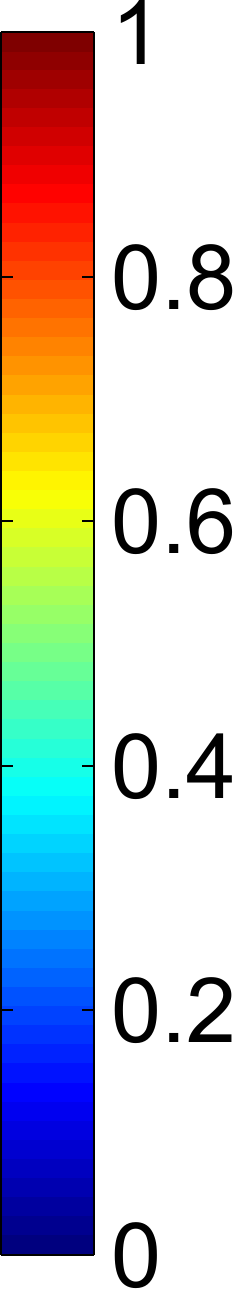}
    
	\end{subfigure} 
	
			\vspace{-0.4cm}
	\caption{Histograms of requests %processed in fog nodes (a), and those rejected (b)
	at 1.3~GFLOP/(s$\cdot$W) cloud efficiency. Results of EEFFRA.}
	\label{fig:hist2D131130}
	\vspace{-0.5cm}
\end{figure}

\end{onecolumn}

\begin{onecolumn}
\begin{figure}
	\centering
	\includegraphics[width=0.49\textwidth]{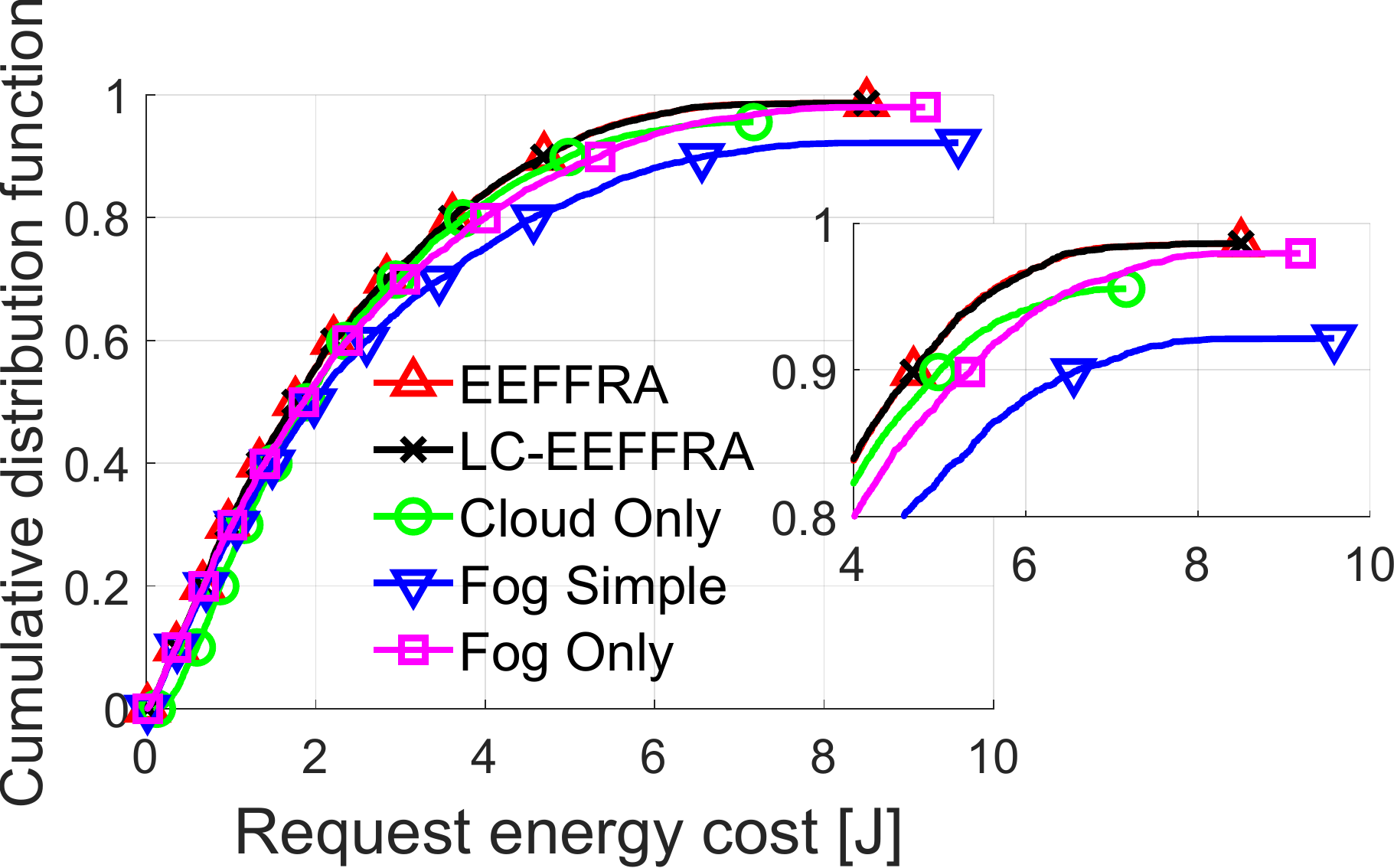}
	\vspace{-0.3cm}
	\caption{\acp{CDF} of request processing energy cost at cloud efficiency of 1.3 GFLOP/(s$\cdot$W). Comparison of different policies.
	} 
	\vspace{-0.3cm}
	\label{fig:cdf_i5}
\end{figure}
\end{onecolumn}
To better illustrate differences in requests allocation policies, we plot \ac{CDF} of energy cost spent on a single request.
Energy spent on rejected requests is assumed to be infinite for the purpose of \ac{CDF} plots.
Such results can be seen in \refFig{fig:cdf_i5}. 
First, it is visible that utilization of both fog and cloud tiers of the network yields significantly better results than utilizing nodes in only one tier.
The proposed \ac{EEFFRA} and LC-\ac{EEFFRA} provide the lowest required energy cost for each percentile of the \acp{CDF}.
All methods do not reach 1 on the y-axis, \ie some requests cannot be processed 
within a given delay budget.
Our methods achieve the lowest rejection rate as shown in the inset of \refFig{fig:cdf_i5}.

\vspace{-0.5cm}
\subsection{Impact of Delay Requirements and Size of Requests on the Offloading Decisions}
\label{sec:size_et_al}

Let us see how the energy consumption and the percentage of rejected requests change with the size and the delay requirement of computation requests.
The energy efficiency of the cloud is set to 1.3~GFLOP/(s$\cdot$W) and parameter sweeps for other parameters are performed. 
All other parameters are generated according to \refTab{tab:ranges}.

\refFig{fig:time} plots \acp{CDF} of energy consumption costs of processing single requests with delay requirements: 100~ms (\refFig{fig:time}a) and 200~ms (\refFig{fig:time}b).
At the required delay of 100~ms, all methods have high rejection rates, with \textit{Cloud Only} being clearly the worst-suited for low-latency applications.
The differences between the rest of the methods are minor -- the requests with such low delay requirements either can or cannot be solved in time at the receiving \ac{FN} and the ability of nodes to transmit tasks between themselves does not improve performance.
With a 200~ms, the differences between approaches become more profound.
Utilizing both fog and cloud (\ac{EEFFRA}, LC-\ac{EEFFRA}) gives the lowest rejection rates and energy costs. 
\textit{Fog Simple} meanwhile has the worst performance.

\begin{twocolumn}
\begin{figure}
	\centering

	\begin{subfigure}{.24\textwidth}
		\centering
		\includegraphics[width=0.999\textwidth]{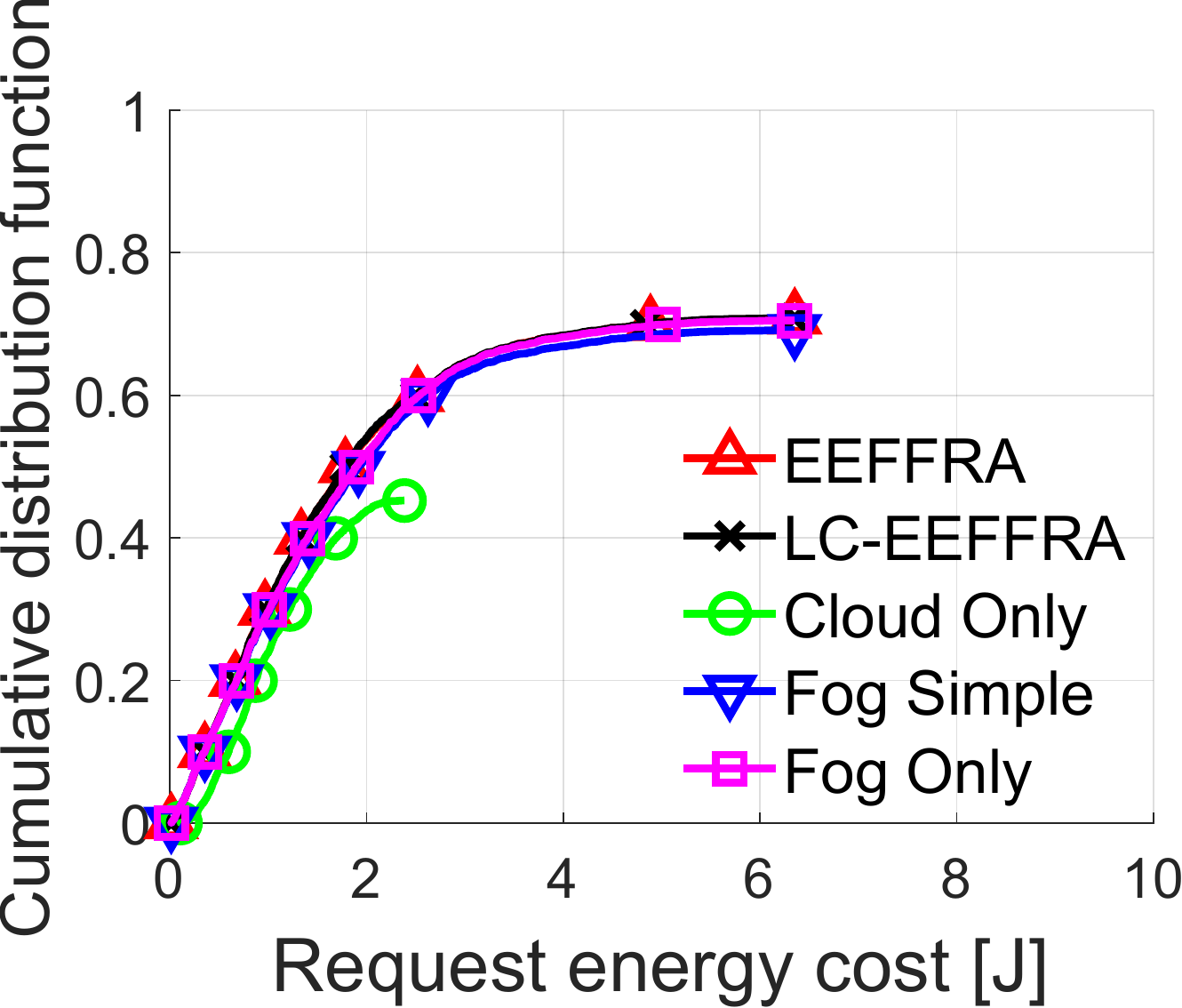}
		\caption{Delay req.: 100~ms.}
	\end{subfigure}
	
		\begin{subfigure}{.24\textwidth}
		\centering
		\includegraphics[width=0.999\textwidth]{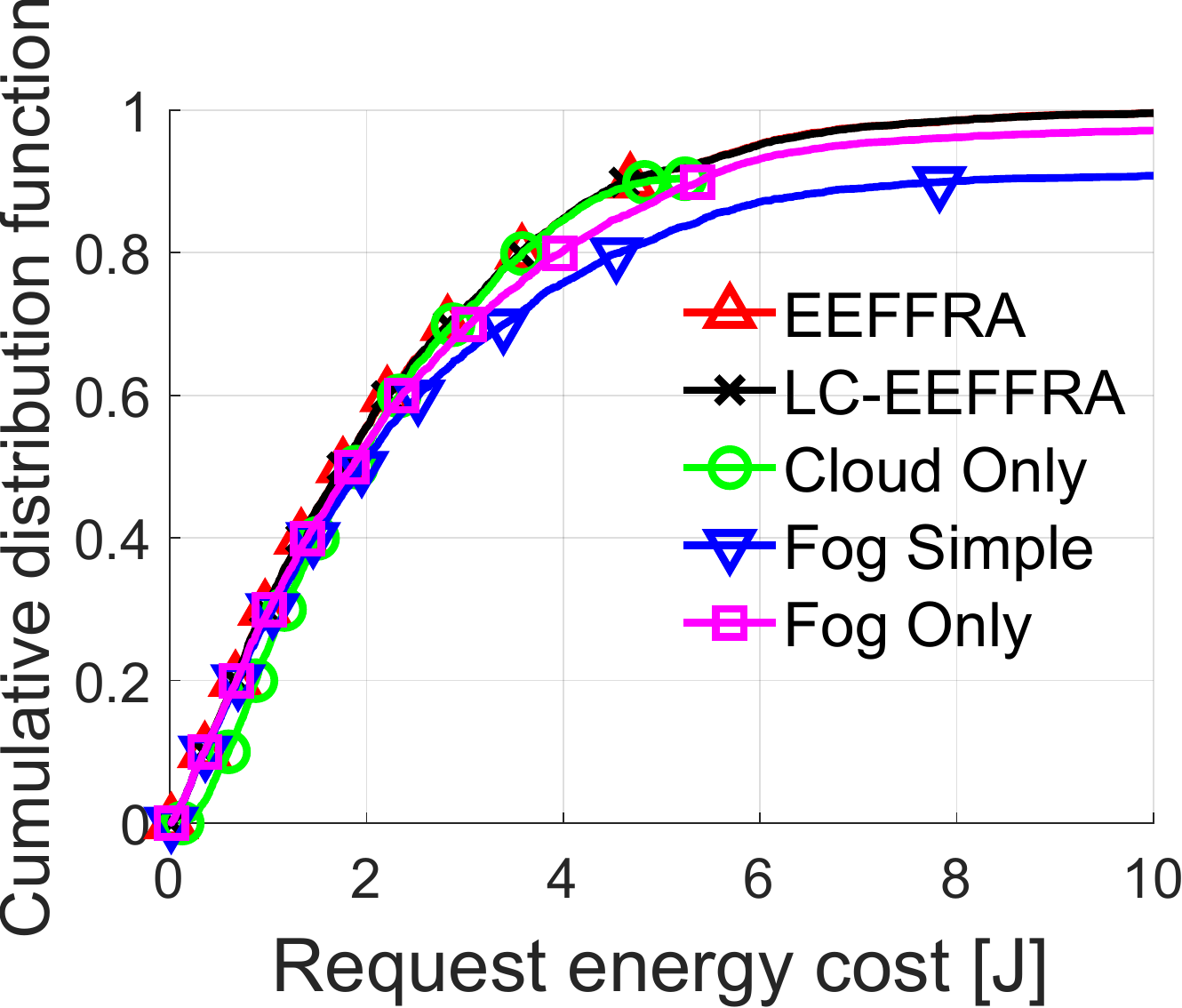}
		\caption{Delay req.: 200~ms.}
	\end{subfigure}%\hfill%
	\begin{subfigure}{.24\textwidth}
		\centering
		\includegraphics[width=0.999\textwidth]{plots/cdfD1000.pdf}
		\caption{Delay req.: 1000~ms.}
	\end{subfigure}
	\caption{\acp{CDF} of request processing energy cost -- influence of delay requirement of requests.} 
	\label{fig:time}
\end{figure}
\end{twocolumn}
\begin{onecolumn}
\begin{figure}
	\centering
	\begin{subfigure}{.35\textwidth}
		\centering
		\includegraphics[width=0.99\textwidth]{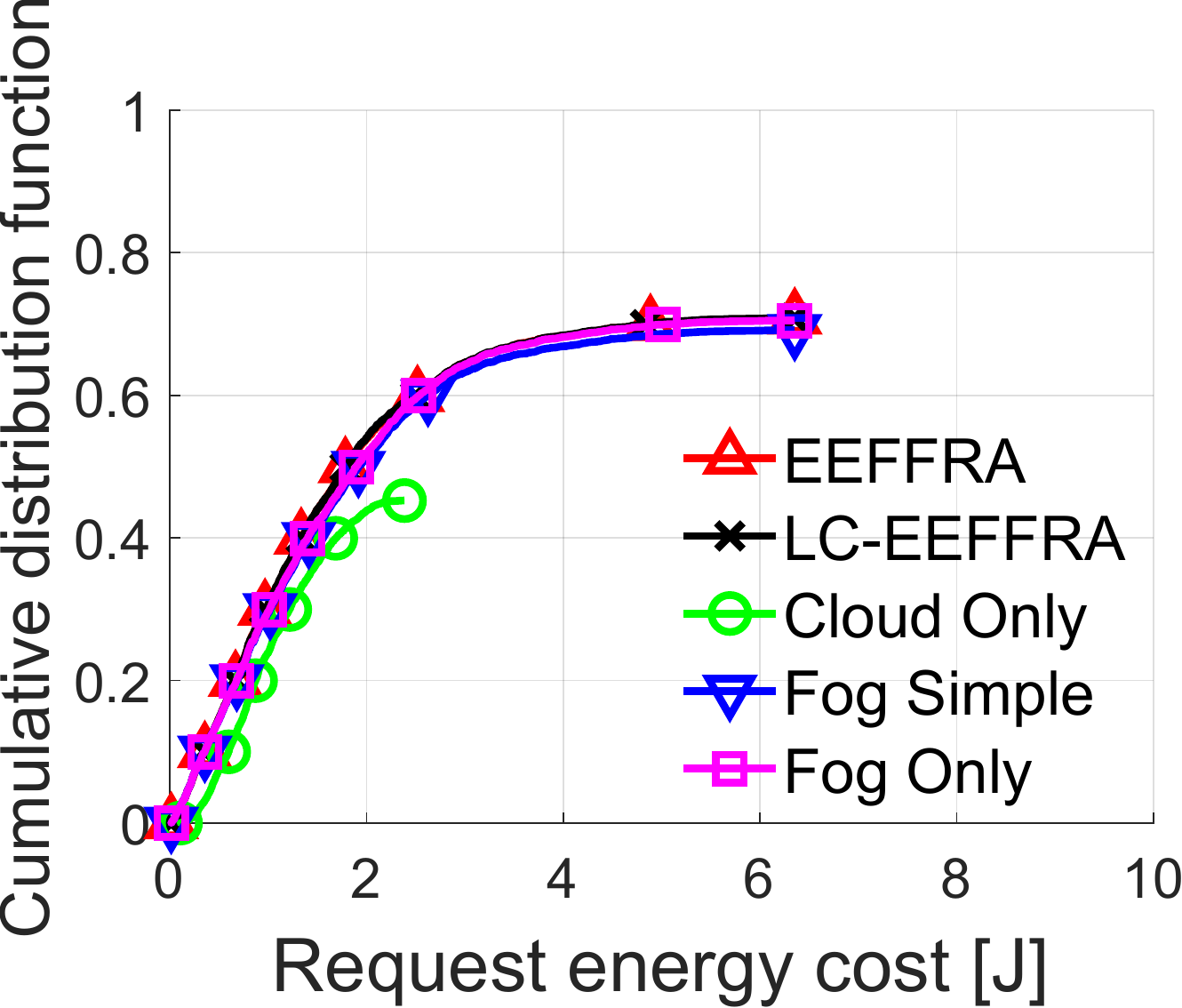}
		\caption{Delay req.: 100~ms}
	\end{subfigure}
	\hspace{0.5mm}
		\begin{subfigure}{.35\textwidth}
		\centering
		\includegraphics[width=0.99\textwidth]{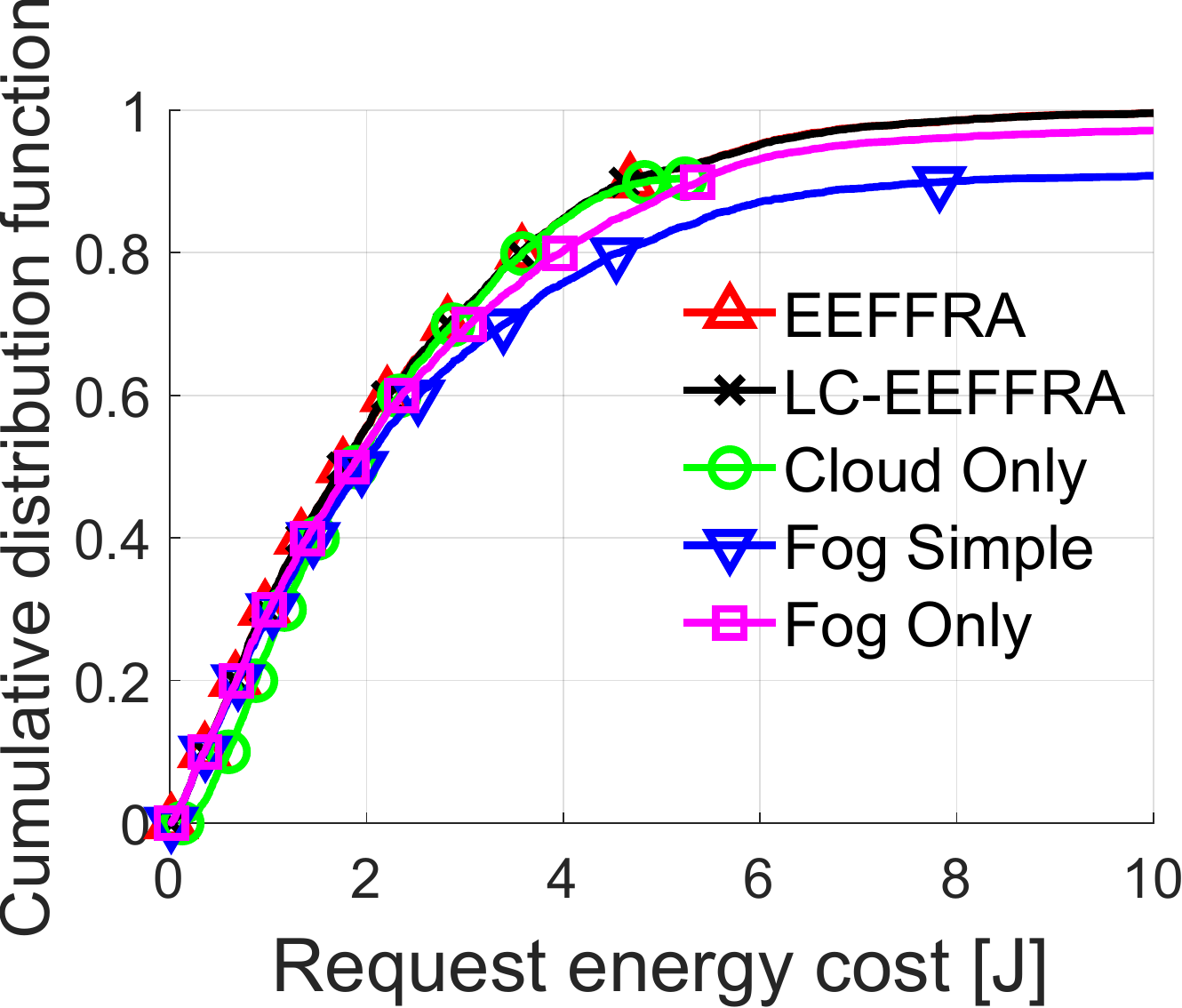}
		\caption{Delay req.: 200~ms}
	\end{subfigure}%\hfill%

	\vspace{-0.4cm}
	\caption{\acp{CDF} of request processing energy cost -- influence of delay requirement of requests.} 

	\label{fig:time}
	\vspace{-0.6cm}
\end{figure}
\end{onecolumn}
% ==================================
% ==================================

\refFig{fig:size} plots \acp{CDF} of energy consumption costs of processing single requests with sizes 5~MB, (\refFig{fig:size}a) and 10~MB (\refFig{fig:size}b).
\ac{EEFFRA}, LC-\ac{EEFFRA}, and \textit{Cloud Only} show the highest energy efficiency for 5~MB requests.
However, \textit{Cloud Only} fails to process a small number of requests within a given delay requirement, while \textit{Fog Only}, \ac{EEFFRA}, and LC-\ac{EEFFRA} process all requests successfully.
\refFig{fig:size}c shows that \ac{EEFFRA} and LC-\ac{EEFFRA} achieve the lowest energy costs and rejection rates for 10~MB requests.
It is worth observing that for both request sizes \textit{Cloud Only} has the highest energy costs up to at least 20-th percentile (caused by energy spent for transmission), but for higher percentiles (influenced by requests with higher arithmetical intensities) its costs are lower than that of either \textit{Fog Only} or \textit{Fog Simple}.
\begin{twocolumn}
\begin{figure}
	\centering

	\begin{subfigure}{.35\textwidth}
	\centering
	\includegraphics[width=0.99\textwidth]{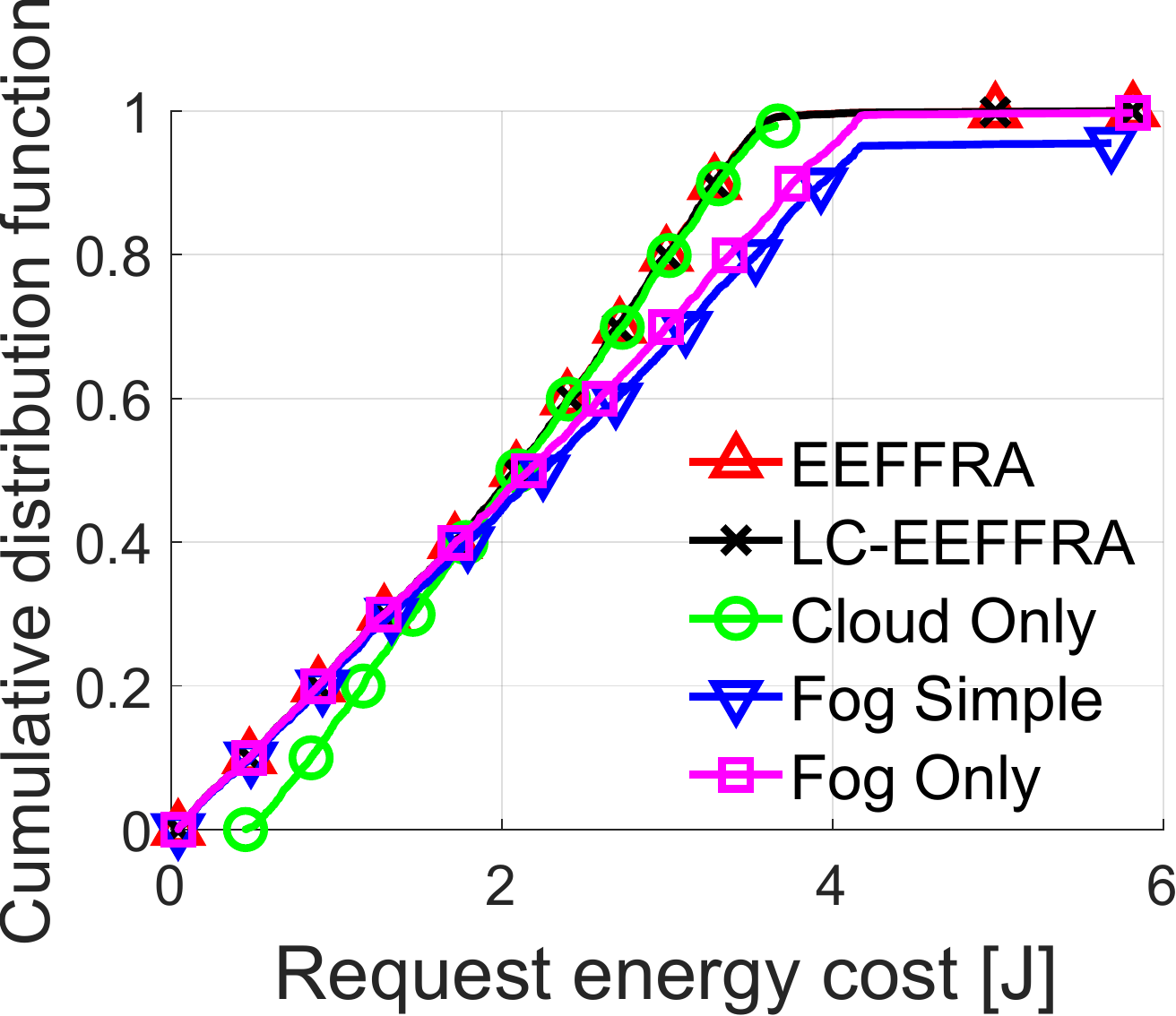}
	\caption{Size: 5~MB% for different %offloading 
%	policies %andcloud efficiency
    \label{fig:size:5MB}
	}

	\end{subfigure}%
	\begin{subfigure}{.35\textwidth}
	\centering
	\includegraphics[width=0.99\textwidth]{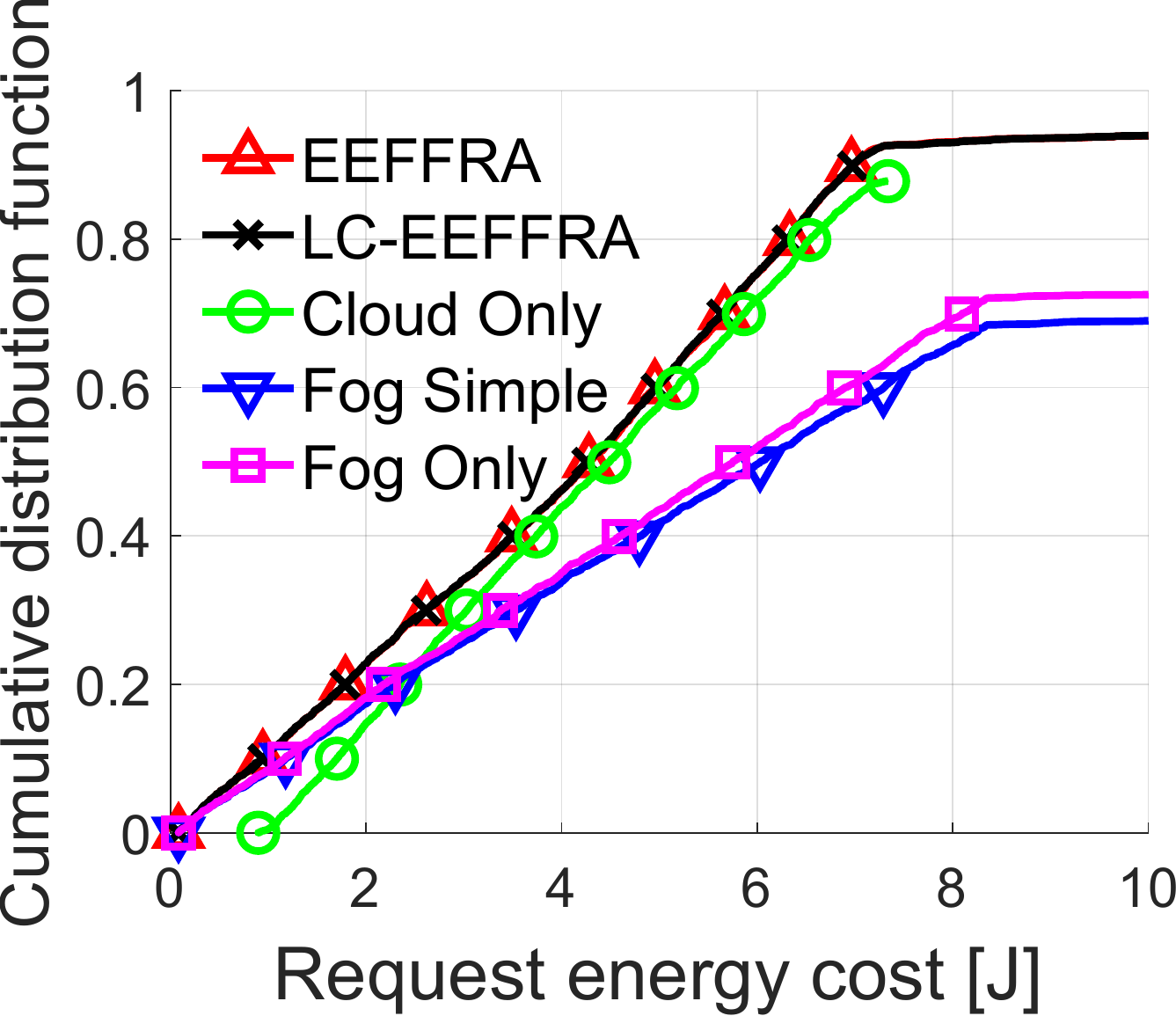}
	\caption{Size: 10~MB% for different %offloading 
%	policies %andcloud efficiency
    \label{fig:size:10MB}
	}

	\end{subfigure}%	\textbf{}
	\caption{\acp{CDF} of request processing energy cost -- influence of size of requests.} %(note different x-axes)
	\label{fig:size}
    \vspace{-3mm}
\end{figure}
\end{twocolumn}
\begin{onecolumn}
\begin{figure}
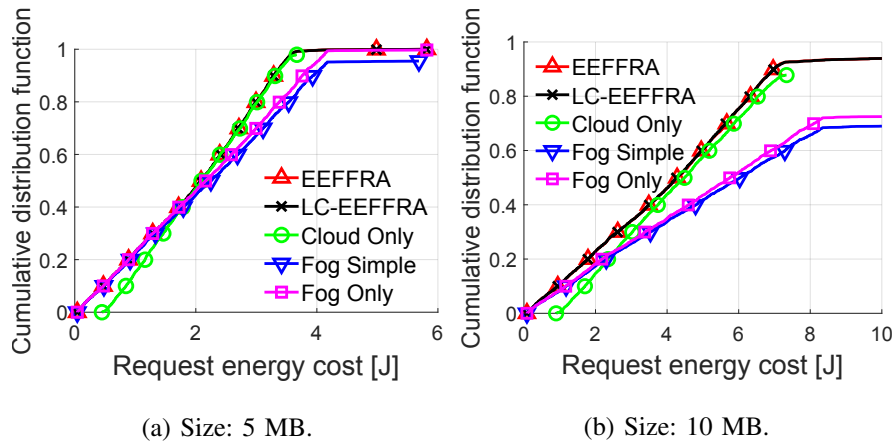

	\centering
	
	\begin{subfigure}{.35\textwidth}
	\centering
	\includegraphics[width=0.99\textwidth]{plots/cdfS5x6.pdf}
	\caption{Size: 5~MB.% for different %offloading 
%	policies %andcloud efficiency
	}

	\end{subfigure}%
	\hspace{0.5mm}
	\begin{subfigure}{.35\textwidth}
	\centering
	\includegraphics[width=0.99\textwidth]{plots/cdfS10x10.pdf}
	\caption{Size: 10~MB.% for different %offloading 
%	policies %andcloud efficiency
	}

	\end{subfigure}%	\textbf{}
	\vspace{-0.4cm}
	\caption{\acp{CDF} of request processing energy cost -- influence of size of requests% (note different x-axes)
	.} 
	\label{fig:size}
    \vspace{-0.5cm}
\end{figure}
\end{onecolumn}
% ==================================

\vspace{-0.5cm}
\subsection{Impact of CPU Frequency of Fog Nodes}

In previous sections, it was assumed that \acp{FN} can dynamically adjust their operating frequency (and voltage) to minimize energy consumption while satisfying delay requirements.
Let us assume that all \acp{FN} utilize the same, fixed CPU frequency. 
\refFig{fig:fixedFreq} shows the average energy cost and percentage of rejected requests plotted as a function of this fixed frequency (swept between 1.6 and 4.2 GHz with a 0.1 GHz step).
Results for \textit{Cloud Only} are constant as no requests are processed in \acp{FN}.
\textit{Fog Simple} and \textit{Fog Only} methods have high rejection rates at low frequencies.
Meanwhile \ac{EEFFRA} and \ac{LC}-\ac{EEFFRA} have the lowest rejection rates while also having the lowest (considering the rejected requests are not taken into account by this metric) energy costs.
As the frequency of \acp{FN} increases the rejection rates decline and average energy cost increases for all methods utilizing \acp{FN}.
However, this effect is considerably smaller for \ac{EEFFRA} and \ac{LC}-\ac{EEFFRA} (utilizing resources in both fog and cloud tiers) than for \textit{Fog Simple} and \textit{Fog Only}.

\begin{figure}
	\centering
	\begin{subfigure}[b]{.35\textwidth}
	\includegraphics[width=0.99\textwidth]{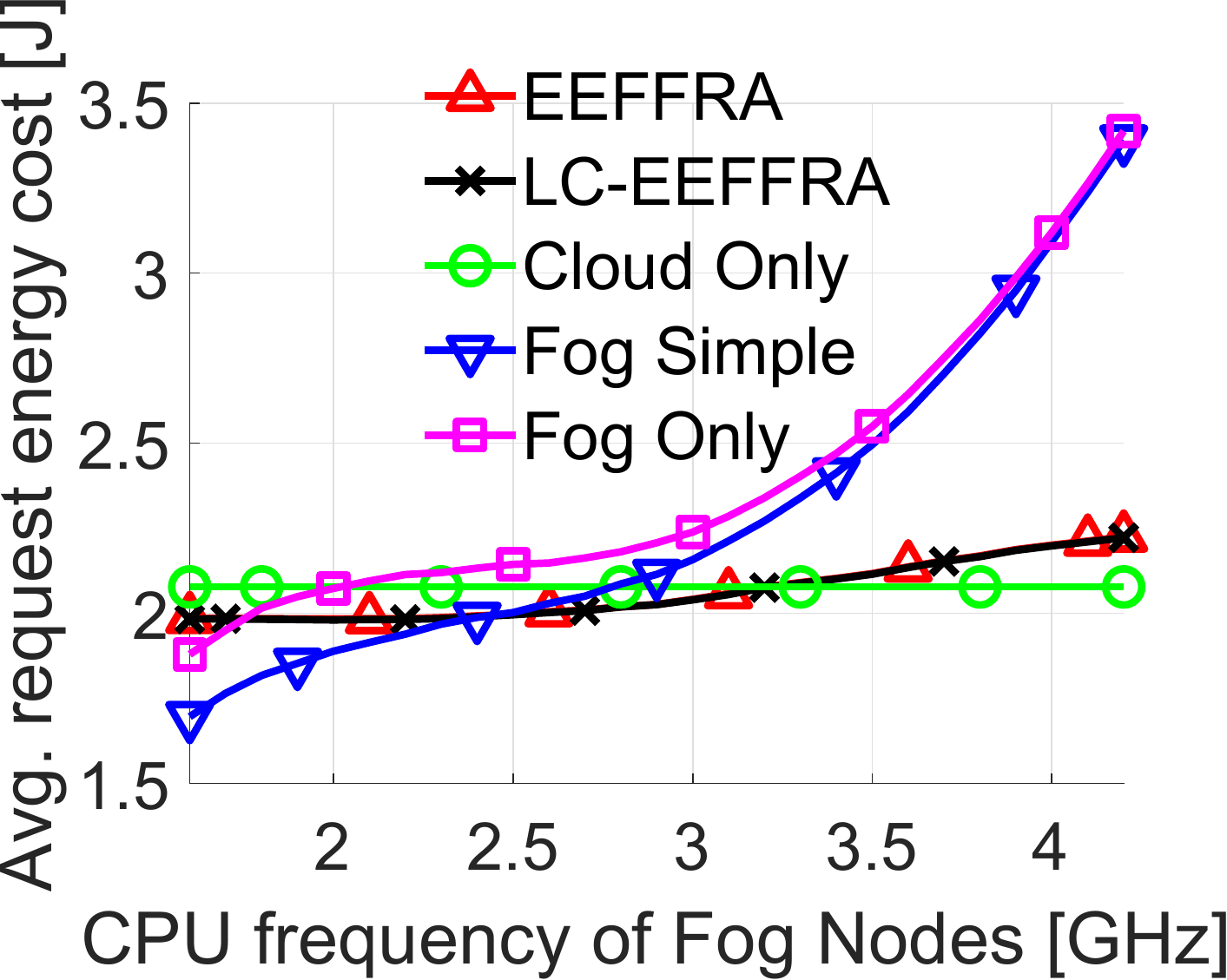}
	\caption{Average energy cost.}
	%spent to process one request for different %offloading 
	%policies}
	\end{subfigure}%\hfill%	\textbf{}
		\hspace{5 mm}
	\begin{subfigure}[b]{.35\textwidth}
	\centering
	\includegraphics[width=0.99\textwidth]{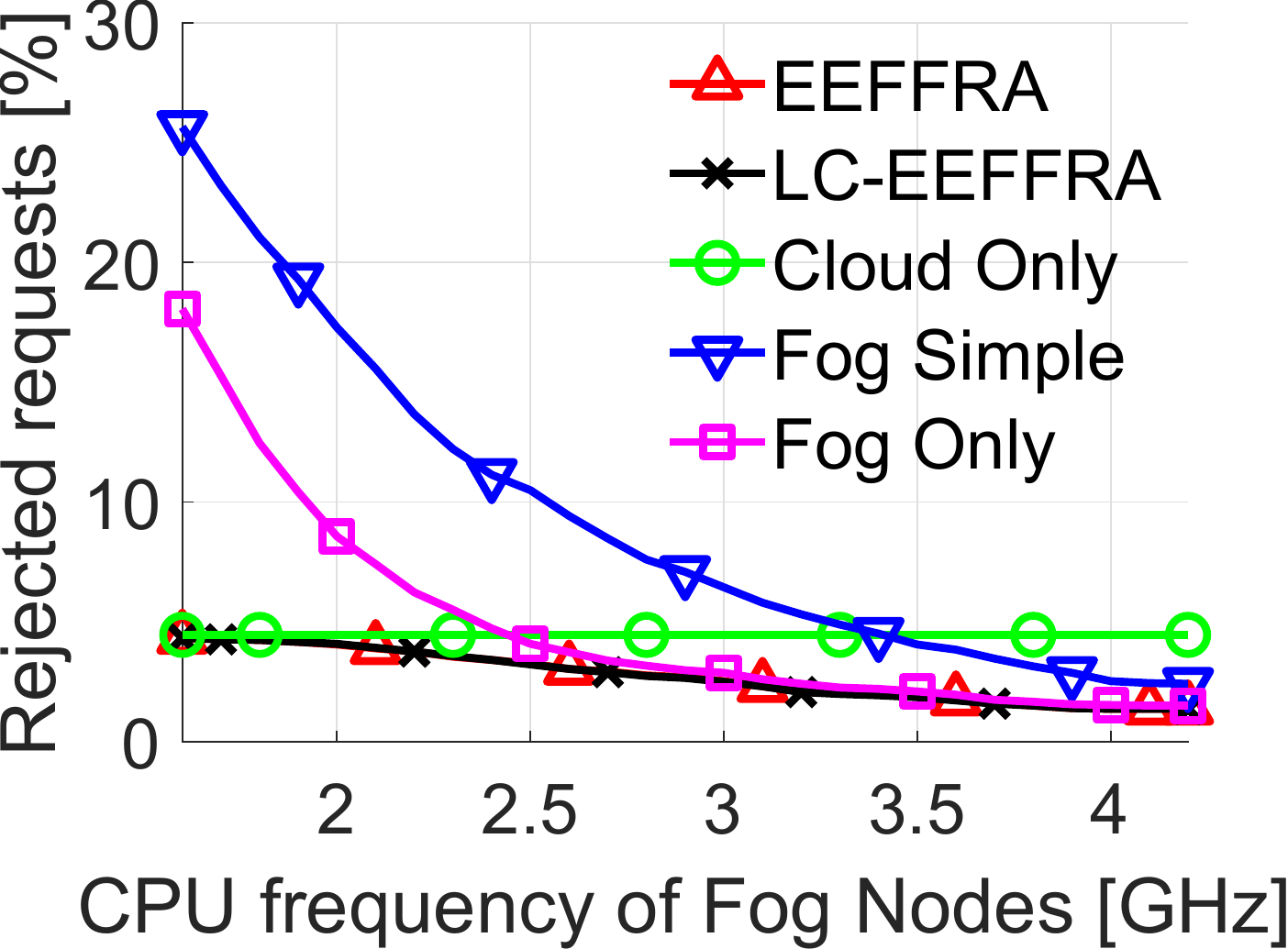}
	\caption{Rejected requests.}% for different %offloading 
	%policies }
	\end{subfigure}%	\textbf{}
	\vspace{-0.4cm}
	\caption{Influence of fixed CPU frequency of FNs.} 
		\vspace{-1.0cm}
	%Avg. energy spent to process one request and percentage of rejected requests
	%-- \textit{base i5 with losses, parameter sweep: arithmetical intensity of requests} }
	\label{fig:fixedFreq}
\end{figure}

Let us compare the efficiency of the network employing \ac{EEFFRA} with and without \ac{DVFS}.
As shown in \refFig{fig:fixedFreq} the possibility to send requests to the cloud diminishes the impact of FNs' operating frequency on energy costs and rejection rate.
Therefore, to focus on the differences, \refFig{fig:cdf_freq_comp1} shows the results of simulations for a network with 10 \acp{FN} and no connection to the cloud.
We compare \acp{CDF} of energy costs per request achieved utilizing \ac{DVFS} with the following fixed frequencies of \acp{FN}: 1.6~GHz (minimum), 2.6063~GHz (optimal frequency for maximizing energy efficiency as seen in \refFig{fig:beta_power_new}, later referred to as 2.6~GHz), and 4.2~GHz (maximum).
The range of possible arithmetic intensities of requests is increased to [1,~500] to make the requests highly variable in terms of required computations speed while the mean time between sets of requests is increased ($\overline{\tk - T_{k-1}}$ = 500~ms).
Rejection rates are increasing with decreasing fixed \ac{FN} frequency.
On the other hand, 4.2~GHz has the highest energy cost.
\ac{EEFFRA} utilizing \ac{DVFS} manages to maintain the %lowest rejection rate and the 
lowest energy cost for every percentile.

\begin{twocolumn}
\begin{figure}
	\centering
		\begin{subfigure}{.24\textwidth}
	\includegraphics[width=0.999\textwidth]{plots/cdfNoCloudInt002long.pdf}
	\end{subfigure}%\hfill%	\textbf{}
		\begin{subfigure}{.24\textwidth}
	\includegraphics[width=0.999\textwidth]{plots/cdfNoCloudInt0002long.pdf}
	
\caption{Traffic: $2~\text{s}^{-1}$
	@ [1,~100]~FLOP/bit.}

	\end{subfigure}%\hfill%	\textbf{}
	
	%\label{fig:cdf_freq_comp1}

	\centering
		\begin{subfigure}{.24\textwidth}
	\includegraphics[width=0.999\textwidth]{plots/cdfNoCloudInt0002longC1to200.pdf}
	\end{subfigure}%\hfill%	\textbf{}
		\begin{subfigure}{.24\textwidth}
	\includegraphics[width=0.999\textwidth]{plots/cdfNoCloudInt0002longC1to500.pdf}
	\end{subfigure}%\hfill%	\textbf{}
	\caption{\acp{CDF} of request processing energy cost -- comparison of \acp{FN} working at fixed frequencies and utilizing \ac{DVFS} (note: different x-axes).}
		\label{fig:cdf_freq_comp1}

	\end{figure}

\end{twocolumn}
\begin{onecolumn}
\begin{figure}
	\centering

	\includegraphics[width=0.35\textwidth]{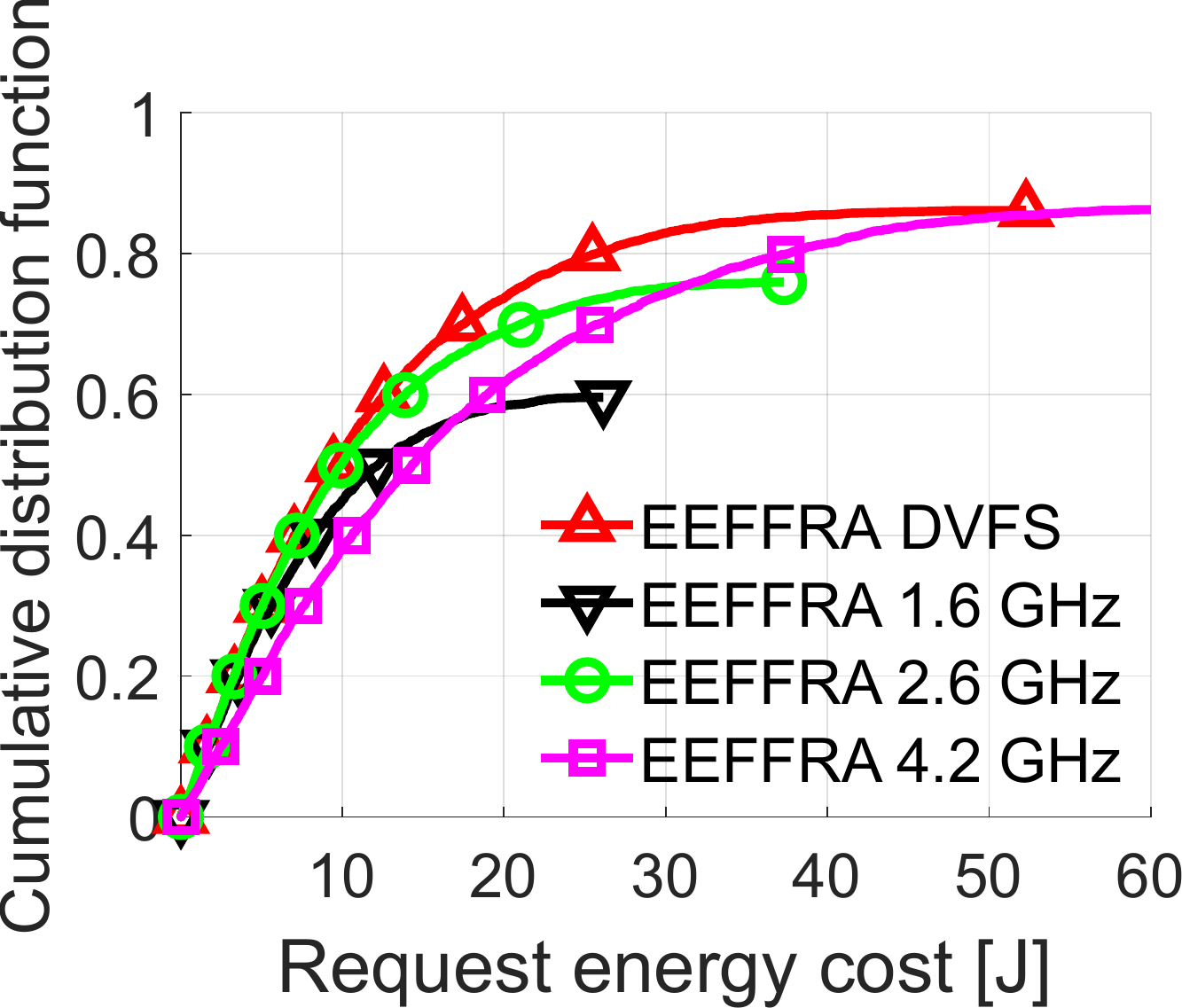}
	\vspace{-0.2cm}
	\caption{\acp{CDF} of request processing energy cost -- comparison of \acp{FN} working at fixed frequencies and utilizing \ac{DVFS}. Parameters: $\overline{\tk - T_{k-1}}$ = 500~ms, $\thetak \in [1,500]$ FLOP/bit.}
		\label{fig:cdf_freq_comp1}
\vspace{-1cm}
	\end{figure}

\end{onecolumn}

% !TEX root = main.tex

\section{Conclusion}
\label{sec:conclusion}
%\acresetall
We have formulated the optimization problem of minimizing the energy consumption in the fog computing network while maintaining the latency constraints. This energy consumption is assumed to be resulting from both transmission and processing of offloaded computational tasks (computation requests originating from the end-users). 
The latency and energy consumption models and their parameters are based on real-life computing and networking equipment product data sheets and measurements.
The proposed \ac{EEFFRA} algorithm solves the proposed optimization problem  using its successive approximations for adjusting clock frequencies of \acp{CPU} in fog nodes.
A~sub-optimal, lower complexity solution \ac{LC}-\ac{EEFFRA}, which does not require coordinated decision making, is also proposed.  
Results of simulations show that both \ac{EEFFRA} and \ac{LC}-\ac{EEFFRA} can significantly reduce  average energy cost and the number of rejected computational requests by distributing the workload between fog and cloud nodes.
The~proposed solutions can be seen as promising alternatives for managing fog computing networks.

%\section*{Acknowledgment}

%\bibliographystyle{IEEEtran}
\bibliographystyle{IEEEtranTCOM}

\bibliography{faust}

\end{document}